\documentclass[aps,pre,twocolumn,amsmath,amsfonts,amssymb,floatfix,superscriptaddress,nofootinbib]{revtex4}
\usepackage{epsfig,amsmath,amssymb,graphics,color,calc}
\usepackage{mathtools}

\def\to{\rightarrow}

\allowdisplaybreaks

\DeclarePairedDelimiter\ket{\lvert}{\rangle}
\DeclarePairedDelimiterX\braket[2]{\langle}{\rangle}{#1 \delimsize\vert #2}

\begin{document}

\title{Delocalization and ergodicity of the Anderson model on Bethe lattices}

\author{Giulio Biroli}
\affiliation{Laboratoire de Physique Statistique, Ecole Normale Sup\'erieure,
PSL Research University, 24 rue Lhomond, 75005 Paris, France}

\author{Marco Tarzia} 
\affiliation{LPTMC, CNRS-UMR 7600, Sorbonne Universit\'e, 4 Pl. Jussieu, F-75005 Paris, France}


\begin{abstract}
In this paper we review the state of the art on the delocalized non-ergodic regime of the Anderson model on Bethe lattices. We also present new results using Belief Propagation, 
which consists in solving the self-consistent recursion relations for the Green's functions directly on a given sample. This allows us to numerically study very large system sizes and to directly access observables related to the eigenfunctions and energy level statistics, such as level compressibility and eigenstates correlation functions. In agreement with recent works,
we establish the existence of a delocalized non-ergodic phase on Cayley trees. On random regular graphs instead 
our results indicate that ergodicity is recovered when the system size is larger than a cross-over scale $N_c (W)$, which diverges exponentially fast approaching the localization transition. This scale corresponds to the 
size at which the mean-level spacing becomes smaller than the Thouless energy $E_{Th} (W)$.
Such energy scale, which vanishes exponentially fast approaching the localization transition, is the one  below which ergodicity in the level statistics is restored in the thermodynamic limit. Remarkably, the behavior of random regular graphs below $N_c (W)$ coincides with the one found close to the root of loop-less infinite Cayley trees, {\it i.e.} only above $N_c (W)$ the effects of loops emerge and random regular graphs behave differently from 
Cayley trees. \\
All in all, our results indicate that ergodicity is recovered in the thermodynamic limit on random regular graph. This notwithstanding, all observables probing volumes smaller than $N_c(W)$ and times smaller than $\hbar/E_{Th} (W)$ are expected to behave as if there were an intermediate phase. Given the very fast divergence of $N_c(W)$ and $\hbar/E_{Th} (W)$ these non-ergodic effects are very pronounced in a large region preceding the localization transition, and they can be related to the intermediate phase present on Cayley trees. 
    
\end{abstract}


\maketitle


\section{Introduction}

After more than a half century, the subject of Anderson localization is still very much alive~\cite{fiftylocalization} as proved by the recent
observations of Anderson localization of atomic gases in one dimension~\cite{aspect} and of classical sound elastic waves
in three dimensions~\cite{localizationelastic}. On the theoretical side several questions remain open: Although there is by now a good
understanding of the localization transition in low dimensional systems, culminating in a functional renormalization group analysis by
a $2+\epsilon$ expansion~\cite{ludwig}, the behavior in high dimensions~\cite{largeD}, in particular the existence of an upper
critical dimension and the relationship with Bethe lattice analysis~\cite{abou}, is still an issue.
Recently, there has been a renewal of interest on
this problem because of its relationship with Many-Body
localization (MBL)~\cite{BAA}, a fascinating new kind of phase transition between a
low temperature non-ergodic phase---a purely quantum
glass---and a high temperature ergodic phase.
This phenomenon has been argued to take place for several
disordered isolated interacting quantum systems, in particular
disordered electrons~\cite{BAA}, and was also independently investigated
in~\cite{wolynes} to explain the quantum ergodicity
transition of complex molecules.
MBL can be thought of as localization in
the Fock space of Slater determinants, which play the
role of lattice sites in a disordered Anderson tight-binding
model.
A paradigmatic representation of this transition~\cite{A97,BAA,jacquod,wolynes,scardicchioMB} 
is indeed (single-particle) Anderson localization
on a very high dimensional lattice, which for spinless electrons consists in an $N$-dimensional hyper-cube 
(where $N\gg1$ is the number of sites of the lattice system). Anderson localization on Cayley tress and Bethe lattices is a drastic simplification of this problem. It is very useful to obtain a qualitative understanding but   
neglect correlations between energies and rare loops. 

Localization had an impact on several fields, in particular Random Matrices and Quantum Chaos. 
As a matter of fact, in the delocalized phase the level statistics is described by random matrix theory and generally corresponds 
to the Gaussian Orthogonal Ensemble (GOE), whereas instead
in the localized phase is determined by Poisson statistics because wave-functions close in energy are exponentially 
localized on very distant sites and hence do not overlap; thus, contrary to the GOE case, there is no level-repulsion and eigen-energies 
are distributed similarly to random points thrown on a line. 

The relationship with quantum chaos goes back to the Bohigas-Giannoni-Schmidt conjecture, which states that the level statistics 
of chaotic (or ergodic) systems is given by random matrix theory, whereas integrable systems instead are characterized by Poisson
statistics~\cite{BGS}. This result can be fully worked out and understood in the semi-classical limit~\cite{berry,altshulerchaos}: for a quantum chaotic
system, in the $\hbar \rightarrow 0$ limit, wave-functions at a given energy become uniformly spread over the micro-canonical 
hyper-surface of the configuration space. They are fully delocalized as expected for an ergodic classical system that covers 
regions with same energy uniformly. Instead, quantum non-ergodic models, such as integrable systems,
are characterized by Poisson statistics and localized wave-functions. All those results support a general
relationship between delocalization--GOE statistics--ergodicity (similarly between localization--Poisson statistics--lack of ergodicity).

However, recent numerical studies~\cite{noi,scardicchio1,scardicchio2,ioffe1,ioffe2,ioffe3} 
of the Anderson model on a Random-Regular Graph (RRG)---a random lattice that has locally a tree-like structure but does not
have a boundary, see below for a precise definition---seem to indicate the possibility of the
existence of a novel intermediate delocalized but non-ergodic phase in a broad disorder range, as first suggested in~\cite{A97}. 
Such phase should be characterized by multifractal eigenfunctions (with the fractal dimensions depending on the disorder strength), anomalous (sub-diffusive) transport along
rare, ramified, paths, and, possibly, non-universal level statistics on a scale larger than the mean-level spacing (while the level statistics on the scale of the mean-level spacing 
 is expected to be described by the GOE ensemble). 
The arguments in favour of this scenario rely mostly on numerical results obtained from Exact Diagonalization (ED) of large but finite 
samples~\cite{noi,scardicchio1,scardicchio2,ioffe1} and 
on an analytic approximation scheme based on Replica Symmetry Breaking (RSB) and ``inflationary population dynamics'' 
developed {\it ad hoc} to deal 
with non-ergodic states~\cite{ioffe1,ioffe2,ioffe3}. 

The possibility of a multifractal delocalized phase in a disordered system is clearly very intriguing, 
especially due to its relationship with MBL.
In fact, this scenario is explicitly realized in suitable models which possess critical states 
such as the Rosenzweig-Porter random matrix~\cite{kravtsov,facoetti} and the power-law random banded matrix~\cite{PLRBM} models, and also occurs
in the tight-binding Anderson model on the (loop-less) Cayley tree, as recently shown in~\cite{mirlin_cayley,garel}.
However, it appears to be in explicit conflict with the analytical 
predictions 
based on the supersymmetric approach for the Anderson model on sparse random graphs~\cite{SUSY,fyod}.
In fact the supersymmetric analysis indicates that the Inverse Participation Ratio (IPR) 
defined as $\Upsilon_2 = \langle \sum_{i=1}^N
|\braket{i}{m}|^4 \rangle$ (where $\braket{i}{m}$ is the amplitude of the 
wave-function $\ket{m}$ on site $i$), 
scales as $\Upsilon_2 \sim C/N$ 
(where the prefactor $C$ depends on the disorder strength, approaching its Gaussian-ensemble value 3 deeply in the metallic 
phase and diverging as $\ln C \sim (W_L - W )^{-1/2}$ at the localization transition).

Moreover, recent numerical investigations 
based on the finite-size scaling of energy levels and wave-functions statistics on the
delocalized side of the Anderson model on RRG~\cite{mirlin} and similar sparse random lattices~\cite{levy,lemarie} 
provided new indications against the existence of a truly intermediate 
non-ergodic extended phase.
Such indications rely on the observation of a non-monotonous behavior of the observables as a function
of the system size on the delocalized side of the transition, which can be explained in terms of (i) the presence of a characteristic
scale which diverges exponentially fast approaching the transition and is already very
large far from it; (ii) the localized nature of the critical point in the limit of infinite 
dimension~\cite{largeD,fyod,efetov,Zirn}.
The combination of these two elements are argued to 
produce dramatic and highly non-trivial finite size effects even very far from
the critical point, and give rise to a strong non-ergodic behavior in a crossover region where the correlation volume $N_c(W)$
is larger than the accessible system sizes. Still, important questions remain answered. 
Probably, the most puzzling feature is the fact that the non-ergodic crossover region observed when the system size is smaller than the correlation volume exhibits non-trivial disorder-dependent (apparent) fractal exponents associated to the spectral statistics, which are independent on $N$ in a broad range of system sizes smaller  than $N_c$. Note that, strictly speaking, these exponents are not rigorously defined since the system is ergodic in the thermodynamic limit. However, since $N_c(W)$ is so large, an effective non-ergodic behavior, that one can describe with effective exponents on several decades, is observed. The main questions are then: What gives rise to this 
effective non-ergodic behavior? Why do the effective exponents change with $W$ (usual finite size scaling would imply a behavior independent of $W$ when $N\ll N_c(W)$) ? How can one explain theoretically these phenomena?  

The existence of this controversy, and the fact that several questions remain open in, after-all, a very old model,  is somewhat surprising, especially if one thinks that the
Anderson transition on tree-like lattices allows, in principle, for an exact solution~\cite{abou,SUSY,fyod,efetov,Zirn,ourselves,aizenmann,semerjian}. This can be
obtained in terms of the
self-consistent equations for the Green's functions, which allow to establish the transition point and the
corresponding critical behavior.
Nevertheless, such exact solution is obtained in the limit of infinite system size, and by introducing
an infinitesimal imaginary regulator $\eta$ which gives an infinitesimal broadening to the energy levels, and
which must be sent to zero {\it after} the limit $N \to \infty$. There is a class of important
observables---including the statistics of eigenfunctions and energy levels---which simply cannot be defined 
on infinite lattices: The mere formulation of statistics of normalized extended wave-functions in a closed system
requires the understanding of the thermodynamic limit of finite-size instances. 
In consequence, in order to address these questions one has either to study large but finite system or 
to work on the simulteneous limit $N \to \infty$, $\eta \to 0$, $N^\phi \eta = {\rm cst}$. 
This motivated the authors of Refs.~\cite{ioffe1,ioffe2,ioffe3} to put forward 
the ``inflationary population dynamics'' approximation scheme
mentioned above to deal with this situation.

In this paper we propose a novel approach to study the Anderson model on Bethe
lattices (both RRGs and loop-less Cayley trees). This strategy consists in finding the solution of the self-consistent recursion
relations for the Green's functions directly on random instance of large {\it but finite} sizes.
This approach is well-known both in statistical physics and computer science, and more precisely, 
in the context of spin-glasses and combinatorial optimization problems, and goes under the
name of ``Belief Propagation'' (BP) or ``Message Passing''~\cite{mezard}, and is generically believed to provide 
an accurate and robust approximation. (The BP approach is in fact exact on the Cayley tree, due to the absence of loops,
and is commonly assumed to become asymptotically exact in the $N \to \infty$ limit on the RRG in most 
cases, see~\cite{mezard} and Refs. therein.)
The advantages of the approach presented here are threefold:
First, the BP solution can be found in a linear time in $N$, thereby allowing to investigate sizes of several order of magnitude larger than those currently accessible by ED and to overcome finite size effects even deep-inside the intermediate non-ergodic crossover regime.
Second, it allows to unveil the difference between the RRG and the Cayley tree:
		although the self-consistent equations are locally the same, the BP approach is 
		sensitive to the existence of boundary and/or loops, and hence gives 
		substantially different solutions for the two types of lattices.
		Third, it allows, to probe the statistics of energy levels and wave-functions' coefficients. In particular, we analyze the level compressibility $\chi_N(E)$ and
the overlap correlation function $K_2 (E)$, which display different scaling behaviors for the ergodic, localized, and multifractal states~\cite{metha,Alts_chi,chalker_chi,mirlin_rev,Bogo,metz,thouless,chalker_K2,krav_K2,kravtsov}.  

The main conclusions of our analysis support the idea that the Anderson model on
the RRG is fully ergodic in the whole delocalized phase (in agreement with~\cite{mirlin,lemarie}). 
Ergodicity is restored on a crossover size $N_c(W)$
which becomes exponentially 
large as the localization transition is approached~\cite{fyod}.
Conversely, we find a genuine non-ergodic extended phase in the Anderson model on the Cayley tree, 
as previously observed in~\cite{garel} and recently predicted in~\cite{mirlin_cayley}.
Interestingly, we show that the non-ergodic features of the apparent intermediate mixed phase observed on the RRG
for system sizes smaller than the correlation volume $N_c(W)$ are essentially controlled by the multifractality of Cayley tree
at the same disorder strength and sufficiently far from the boundary.

The paper is organized as follows.
In the next section we introduce the model and briefly review previous results and studies.
In Sec.~\ref{sec:numerics} we present a detailed and accurate numerical analysis of eigenvalues and eigenvectors statistics 
obtained from ED of the Anderson model on the RRG.
In Sec.~\ref{sec:BP} we show the results of the BP approach for the Anderson model on the RRG and on Cayley trees, 
and highlight the difference between the two kinds of lattices.
Finally, in Sec.~\ref{sec:conclusions} we discuss the physical implications of our results, providing some concluding remarks and perspectives for future work.
Some technical aspects are discussed in details in the appendices~\ref{app:multi}-\ref{app:NE}. 

\section{Model and State of the Art} \label{sec:model}

The model we focus on consists in non-interacting spinless electrons in a
disordered potential: 
\begin{equation} \label{eq:H}
{\cal H} = - t \sum_{\langle i,j \rangle} \left( c_i^{\dagger} c_j
+ c_j^{\dagger} c_i \right ) - \sum_{i=1}^N \epsilon_i \, c_i^\dagger c_i \, ,
\end{equation}
where the first sum runs over all the nearest neighbors sites of the
lattice, the second sum runs over all $N$ sites;
$c_i^\dagger$, $c_i$ are fermionic creation and annihilation operators,
and $t$ is the hopping kinetic energy scale, which we take
equal to $1$. The on-site energies $\epsilon_i$ are i.i.d. random variables uniformly
distributed in the interval $[-W/2,W/2]$:
\begin{equation} \label{eq:peps}
p(\epsilon) = \frac{1}{W} \, \theta \! \left ( \frac{W}{2} - | \epsilon | \right) \, .
\end{equation}

As anticipated in the introduction, we will focus on two types of Bethe lattices with a tree-like structure.
The first is defined as a $(k+1)$-RRG, {\it i.e.}, a lattice
chosen uniformly at random among all graphs of $N$ sites where each
of the sites has connectivity $k+1$. 
The properties of such random graphs have been extensively studied (see Ref.~\cite{wormald} for a review).
A RRG can be essentially viewed as a finite portion of a tree wrapped onto itself.
It is known in particular that for large number of sites any finite portion of such a graph is a tree with 
a probability going to one as $N \to \infty$, and that the RRG 
has large loops of typical length of order $\ln N$~\cite{wormald}.
Hence the RRG ensemble can be thought as describing a tight-binding model on a lattice that has locally a tree-like structure but does not possess a boundary.
The model~(\ref{eq:H}) is then a sum of two random matrices, ${\cal H} = {\cal C} + {\cal D}$: 
${\cal C}$ is the connectivity matrix of the RRG, ${\cal C}_{ij} = -t$ if sites $i$ and $j$ are connected
and zero otherwise. ${\cal D}$ is the diagonal matrix corresponding to the on-site random energies, 
${\cal D}_{ij} = \epsilon_i \delta_{ij}$.
It is known from previous studies that the former ensemble of sparse random matrices belongs
to the GOE universality class (with fully delocalized eigenvectors)~\cite{RRG-GOE,Bauerschmidt},
while the latter is described by definition by Poisson statistics (with fully localized eigenvectors).

The second type of lattice that we will consider is a (non-random) finite portion of $n_g$ generations 
of an infinite loop-less tree of connectivity $k+1$ (also known as Cayley tree). 
A finite fraction of the sites of a Cayley tree belong to the boundary and only have connectivity equal to $1$ 
(more precisely, for a Cayley tree of $n_g$ generations, the number of boundary sites
is $(k+1)k^{n_g-1}$, while the total number of sites is $1 + (k+1)(k^{n_g}-1)/(k-1)$)~\cite{canopy}.
Note that while the RRG is statistically translationally invariant, the Cayley tree is not translationally invariant
even in absence of disorder, since the properties of a given site depend on its distance from the boundary 
(or, equivalently, from the root) of the tree.

Localization on the RRG was first studied by Abou-Chacra, Anderson
and Thouless~\cite{abou} and then later by many others,
see~\cite{noi,scardicchio1,scardicchio2,ioffe1,ioffe2,ioffe3,mirlin,lemarie,fyod,efetov,Zirn,Verb,berkovits,ourselves,aizenmann,semerjian,metz} and Refs.~therein.
Many similarities, but also few important differences, with the $3d$ behavior have been found.
As mentioned above, the differences mainly concern the critical properties. Contrary to the finite dimensional case, the critical behavior is not power-law-like 
but instead exponential, i.e., one finds essential singularities approaching the localization transition from the delocalized regime~\cite{fyod,efetov,Zirn}. Moreover, the IPR, 
is found to have a discontinuous jump at the
transition from a $O(1)$ toward a $1/N$ scaling~\cite{SUSY}, instead of being continuous at the transition. 
Arguments based on supersymmetric field theory indicate that
the level statistics should display a transition from GOE to
Poisson statistics concomitant with the localization transition~\cite{fyod,SUSY}.
However, the first numerical studies didn't fully support this claim~\cite{noi,berkovits}.
Moreover, the arguments of~\cite{A97} indicates
that the two transitions might actually not coincide.
As discussed above, evidences of an intermediate phase, which is delocalized and yet still not ergodic
were first found in~\cite{noi}. These findings triggered a lot of activity. 
In Refs.~\cite{scardicchio1,scardicchio2}, based on the numerical study of 
the spectrum of fractal dimensions of finite size systems, it was conjectured that the eigenstates are
multifractal in the whole delocalized phase. More recently, 
the authors of Refs.~\cite{ioffe1,ioffe2,ioffe3} 
combined exact diagonalization and semi-analytical calculations to 
claim the existence of the intermediate non-ergodic but delocalized phase 
in a broad disorder strength $W_E < W < W_L$.
Finally, the numerical investigations of Refs.~\cite{mirlin,levy,lemarie} of the level and eigenfunction statistics 
on the delocalized side of the Anderson transition on the RRG and similar sparse random lattices unveiled the
existence of very strong finite size effects with a characteristic crossover scale $N_c(W)$ associated to a pronounced non-monotonous behavior
of the observables as a function of $N$. Such correlation volume is found to diverge exponentially fast
at the Anderson transition, thus possibly explaining the discrepancy between theoretical results and numerics.  The origin of the non-monotonicity has been traced back to the localized 
nature of the Anderson critical point in the limit of infinite dimensions~\cite{largeD,fyod,efetov,Zirn}:
For $N \ll N_c$ the system flows towards the Anderson transition fixed point, whose properties on the RRG 
are analogous to the localized phase, whereas for $N \gg N_c$ the system approaches the $N \to \infty$
ergodic behavior. The conclusion of the investigations of Refs.~\cite{mirlin,levy,lemarie} is thus that
the system is ergodic in the whole delocalized phase, but is characterized by dramatic and non-trivial
finite-size effects even very far from the critical point, which give rise to an apparent non-ergodic
behavoir in a crossover region where the correlation volume is larger than the accessible system sizes.
Nonetheless, some aspects of the problem cannot be explained by this scenario and must be analyzed more carefully. As we stressed in the introduction, important questions on the nature of this cross-over region remain
unswered. 

On the other hand, the properties of the Anderson model on the Cayley tree have been much less studied.
Monthus and Garel studied numerically the statistics of transmisson amplitudes on a Cayley tree, finding
that it has a multifractal form in the delocalized phase~\cite{garel}.
More recently, these results have been confirmed by the analysis of~\cite{mirlin_cayley} 
where it was shown that the delocalized phase have subtle properties and
is, in fact, non-ergodic, with wave-functions presenting a 
multifractal behavior.

In the following, without loss of generality,
we focus on the $k=2$ case ({\it i.e.}, total connectivity $k+1 = 3$) and (mostly) on the middle of the spectrum,
$E=0$. Previous studies of the transmission properties and dissipation propagation determined
that the localization transition takes place at $W_L \approx 18.2$~\cite{abou,garel,ourselves}, while 
previous analysis of the spectral properties have suggested the presence of the non-ergodic delocalised phase 
in the range $10 \approx W_E < W < W_L$~\cite{noi,ioffe1,ioffe2,ioffe3}.

\section{Exact diagonalization on the RRG} \label{sec:numerics}

In order to analyze the statistics of energy levels and wave-functions amplitudes, and clarify its relationship with the localization
transition, we have diagonalized the Hamiltonian~(\ref{eq:H}) on the RRG for several system sizes
$N=2^n$, from $n=6$ to $n=15$, and for several values of the disorder strength $W$ on the delocalized side of the Anderson
transition $W < W_L \approx 18.2$.
For each $N$ and $W$, we have averaged over both the on-site quenched disorder
and on RRG realizations, taking (at least) $2^{22-n}$ 
different samples.
Since we are interested in $E=0$, we only focused on $1/8$ of the eigenstates centered around
the middle of the band (we have checked that taking $1/16$ or $1/32$ of the states does
not alter the results, but yields a poorer statistics).

\subsection{Level statistics}

We have studied the statistics of
level spacings of neighboring eigenvalues: $s_m = E_{m+1} - E_m \ge 0$,
where $E_m$ is the energy of the $m$-th eigenstate in the sample.
In the delocalized regime, level crossings are forbidden. Hence the eigenvalues are strongly correlated and
the level statistics is expected to be described by Random Matrix Theory (more precisely, several results support a general
relationship between delocalization and the Wigner's surmise of the GOE).
Conversely, in the localized phase
wave-functions close in energy are exponentially localized on very distant sites and do not overlap. Thus
there is no level-repulsion and eigenvalues should be distributed similarly to random points thrown on a line
(Poisson statistics).
In order to avoid difficulties
related to the unfolding of the spectrum, we follow~\cite{huse} and measure the ratio of adjacent gaps, 
\[
r_m = \frac{\min \{ s_m, s_{m+1} \}}{\max \{ s_m, s_{m+1} \}} \, ,
\]
and obtain the
probability distribution $\Pi (r)$, which displays a universal form depending on
the level statistics~\cite{huse}. In particular $\Pi (r)$ is expected to converge to its GOE and
Poisson counterpart in the extended and localized regime~\cite{Pr-GOE}, allowing to discriminate between the two phases
as $\langle r \rangle$ changes from $\langle r \rangle_{\rm GOE} \simeq 0.53$ to $\langle r \rangle_P \simeq 0.39$
respectively.

The GOE-Poisson transition can also be captured by correlations between nearby eigenstates such as
the mutual overlap between two subsequent eigenvectors, defined as:
\[
q_m = \sum_{i=1}^N | \braket{i}{m} | | \braket{i}{m+1} | \, .
\]
In the GOE regime the wave-functions amplitudes are i.i.d. Gaussian random variables
of zero mean and variance $1/N$~\cite{porter-thomas}, hence $\langle q \rangle$
converges to $\langle q \rangle_{\rm GOE} = 2/\pi$. 
Conversely in the localized phase two successive eigenvector are typically peaked around very distant sites and do not overlap, and therefore
$\langle q \rangle_{P} \to 0$ for $N \to \infty$.
At first sight this quantity seems to be related to the statistics of wave-functions' coefficients rather than to energy gaps. 
Nonetheless, in all the random matrix models that have been considered in the literature so far, one empirically finds that $\langle q \rangle$ is directly associated 
to the statistics of gaps between neighboring energy levels. Perhaps the best example of that is provided by the generalization of the Rosenzweig-Porter random matrix model 
of~\cite{kravtsov,facoetti}, where there is a whole region of the parameter space where wave-functions are delocalized but multifractal and strongly correlated, while the 
statistics of neighboring gaps is still described by the GOE ensemble. In this case one numerically finds that $\langle q \rangle$ converges to its GOE universal value 
$2/ \pi$ irrespective of the fact that wave-functions amplitudes are not uncorrelated Gaussian random variables of variance $1/N$.

\begin{figure}
 \includegraphics[angle=0,width=0.48\textwidth]{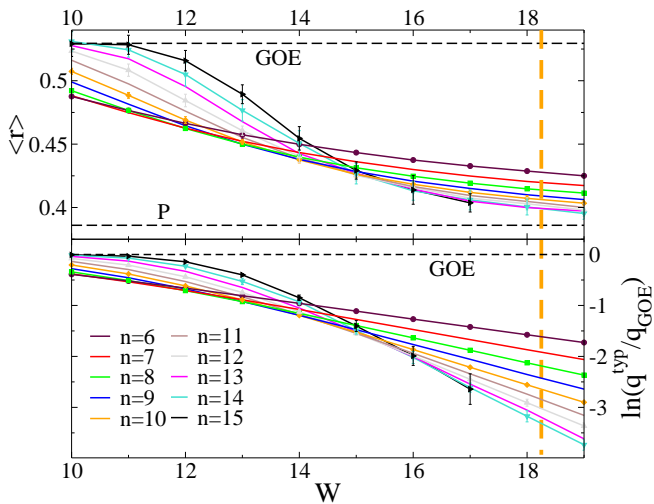}
 \caption{\label{r-q} 
 $\langle r \rangle$ (upper panel) and $\ln (q^{\rm typ}/q_{\rm GOE})$ (lower panel) as a function of the disorder $W$ for several
 system sizes $N=2^n$ with $n$ from $6$ to $15$. The horizontal dashed lines correspond to the reference GOE and Poisson asymptotic values. The vertical 
orange dashed line spots the position of the Anderson localization transition, $W_c \approx 18.2$.}
 \end{figure}
 
 In Fig.~\ref{r-q} we show the behavior of the average value of the ratio of adjacent gaps, $\langle r \rangle$, and
 of (the logarithm of) the typical value of the mutual overlap between subsequent eigenvectors, 
 $q^{\rm typ} = e^{\langle \ln q \rangle}$, 
 as a function of the disorder $W$, for several system sizes $N=2^n$, with $n$ from $6$ to $15$.
As expected, for small (resp. large) enough disorder we recover the universal values
$\langle r \rangle_{\rm GOE} \simeq 0.53$ and $q_{\rm GOE}^{\rm typ} = 2 / \pi$ (resp. $\langle r \rangle_{P} \simeq 0.39$
and $q_P^{\rm typ} \to 0$) corresponding to GOE (resp. Poisson) statistics.
However, as pointed out in~\cite{noi} 
the different curves corresponding to different values of $N$ cross much before the localization transition,
occurring at $W_L \approx 18.2$, as indicated by the vertical dashed line in the plot.
This behavior was 
interpreted in terms of an intermediate delocalized but non-ergodic phase~\cite{noi}.
Nevertheless, analyzing carefully  the data, we realized that the crossing point is in fact slowly but systematically 
drifting towards larger values of $W$ as $N$ is increased, as also observed~\cite{mirlin,levy}. 
  
\begin{figure}
 \includegraphics[angle=0,width=0.46\textwidth]{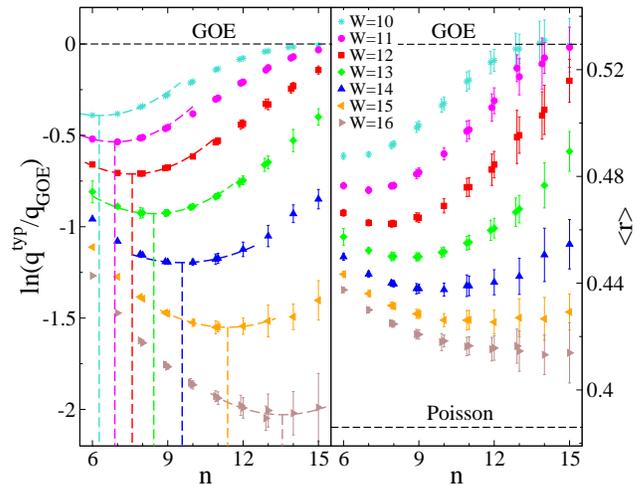}
 \caption{\label{minimo2}
 $\ln (q^{\rm typ}/q_{\rm GOE})$ (left panel) and $\langle r \rangle$ (right panel) as a function of $n=\log_2 N$ for $W=10$ (turquoise stars), 
 $W=11$ (violet circles), $W=12$ (red squares), $W=13$ (green diamonds), $W=14$ (blue up triangles),
 $W=15$ (orange left triangles), and $W=16$ (brown right triangles). The data show the non-monotonic  behavior 
 of $q^{\rm typ}$ and $\langle r \rangle$. The position of the minimum $n_c(W)$ extracted from 
 $q^{\rm typ}(W)$ is represented by the vertical dashed lines.}
 \end{figure}
 
This is clearly unveiled by Fig.~\ref{minimo2}, where we plot the behavior of $q^{\rm typ}$ and $\langle r \rangle$ 
as a function
of $n = \log_2 N$, for several values of the disorder belonging to the range where the curves of
$\langle r \rangle$ and $q^{\rm typ}$ for different $n$ cross, i.e., $10 \lesssim W \lesssim 16$. 
One indeed observes that
in this region
$q^{\rm typ}$ and $\langle r \rangle$ become non-monotonic functions of $n$.
The position of 
the minimum of $q^{\rm typ}$ (highlighted by dashed vertical lines in the left panel of Fig.~\ref{minimo2}) 
naturally defines a characteristic system size, $N_c(W) = 2^{n_c (W)}$, governing the crossover from Poisson to GOE statistics (on the scale of the mean level spacing):
For $N<N_c(W)$ one has indeed that $q^{\rm typ}$ decreases as the system size is increased, as expected for localized 
wave-functions, whereas for $N>N_c(W)$ it is an increasing function of $n$ and eventually converges to the GOE universal 
value.
The same non-monotonic behavior as a function of the system size is found for $\langle r \rangle$ (right panel of Fig.~\ref{minimo2}), and
has been previously observed in Refs.~\cite{ioffe1,mirlin,levy} 

\subsection{Wave-functions statistics: Inverse Participation Ratio, support set, and the spectrum 
of fractal dimensions} \label{sec:IPR-SS}

The IPR of the eigenfunction $\ket{m}$ is defined
as $\Upsilon_{2}^{(m)} = \sum_{i=1}^N | \braket{i}{m} |^4$.
In the full extended regime wave-functions are uniformly spread over all the sites of the RRG,
thus $\braket{i}{m}$ are random variables of order $1/\sqrt{N}$, due to normalization, and 
$\langle \Upsilon_2 \rangle$ vanishes as $1/N$ for $N \to \infty$.
Conversely in the localized phase wave-functions are localized on $O(1)$ sites and 
$\langle \Upsilon_2 \rangle$ approaches a constant value in the thermodynamic limit
(in particular, in the infinite disorder limit one has that $\langle \Upsilon_2 \rangle \to 1$).

A related---and less fluctuating---observable is provided by the support set, recently introduced
in~\cite{scardicchio1,scardicchio2} as a measure of wave-functions ergodicity.
For an eigenvector $\ket{m}$ with sites ordered
according to $|\braket{i}{m}| > |\braket{i+1}{m}|$, it is defined as the sets of sites $i < S_\epsilon^{(m)}$ such that:
\begin{displaymath}
\sum_{i=1}^{S_\epsilon^{(m)}} |\braket{i}{m}|^2 \le 1 - \epsilon < \sum_{i=1}^{S_\epsilon^{(m)}+1} |\braket{i}{m}|^2 \, .
\end{displaymath}
The scaling of $S_\epsilon$ for $N \to \infty$ and $\epsilon$ arbitrary small but finite allows to
discriminate between the extended and the localized regimes, as $S_\epsilon$ is $N$-independent for localized wave-functions while
it diverges as $N$ for $N \to \infty$ for fully delocalized states.

In the intermediate extended non-ergodic 
phase the eigenstates are supposed to be be delocalized on a subset of $N^D$ sites.
One therefore expects that the disorder-dependent fractal exponent $D$ describing the scaling of 
the support set with the system size as $\langle S_\epsilon \rangle \sim N^{D}$
should be strictly smaller than one in the intermediate delocalized non-ergodic phase~\cite{scardicchio1,scardicchio2,ioffe1,ioffe2,ioffe3}.
In fact one can show~\cite{scardicchio1,scardicchio2,ioffe2,ioffe3} that the exponent $D$ coincides with
the fractal dimension $D_1$.
Similarly the IPR should behave as
$\langle \Upsilon_2 \rangle \sim N^{-D_2}$, with $D_2 \in (0,1)$. (See below for a precise definition of the
fractal exponents $D_q$.)

We have measured the wave-functions' amplitudes
from ED of the Hamiltonian~(\ref{eq:H}) on the RRG for several values of the 
disorder strength $W$ and for several system sizes $N = 2^n$, and computed 
the typical value of the IPR, $\Upsilon_2^{\rm typ} = e^{\langle \ln \Upsilon_2 \rangle}$,
and the average value of the support set, $\langle S_\epsilon \rangle$.\footnote{One should focus in the
regime where $\epsilon$ is arbitrary small but finite. In practice we have taken $\epsilon \in (10^{-4}, 10^{-3})$.}
As explained in the previous section, averages are taken over (at least) $2^{22-n}$ different realizations of the disorder and over $1/8$ of the eigenstates
centered around the middle of the band.

\begin{figure}
 \includegraphics[angle=0,width=0.44\textwidth]{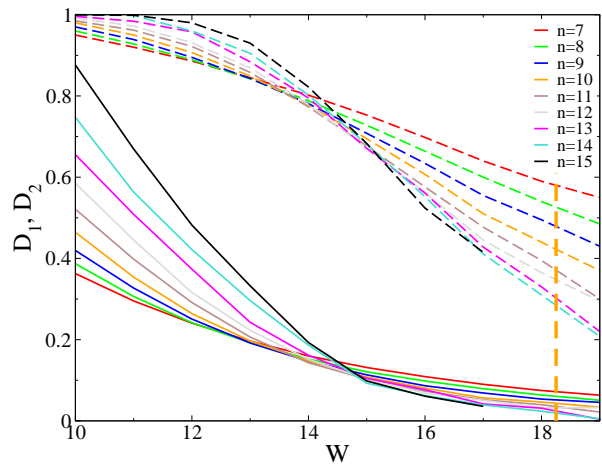}
 \caption{\label{exp-IPR-SS}
Flowing fractal exponents $D_2$ (continuous lines) and $D_1$ (dashed lines) describing the
scaling with the system size of the typical value of the  IPR and of the
average of the support set (see Eq.~(\ref{eq:exp-IPR-SS})) 
as a function of the disorder $W$. Numerical data for different system sizes $N=2^n$ 
are shown, with $n$ from $7$ to $15$. The vertical dashed orange line shows the position of the 
localization transition $W_L$.}
 \end{figure}

 The flowing fractal exponents $D_2$ and $D_1$ describing the scaling of the typical value of the IPR and of the 
average value of the support set with $N$ can then be approximately evaluated as: 
\begin{equation} \label{eq:exp-IPR-SS}
\begin{split}
D_2 (W,n) & \simeq - \, \frac{\ln \Upsilon_2^{typ} (W,n) - \ln \Upsilon_2^{typ} (W,n-1)}{\ln 2} \, , \\
D_1 (W,n) & \simeq \frac{ \ln \langle S_\epsilon (W,n) \rangle - \ln \langle S_\epsilon (W,n-1) \rangle}{\ln 2} \, .
\end{split}
\end{equation}
In Fig.~\ref{exp-IPR-SS} the numerical values of 
$D_2$ and $D_1$ are plotted as a function of the disorder $W$
for several system sizes. 
$D_2$ and $D_1$
show a remarkably similar---although slightly less clean---behavior compared to the one of $\langle r \rangle$ and $q^{typ}$ 
of Figs.~\ref{r-q} and \ref{minimo2}.
At fixed $N$, 
$D_2$ and $D_1$ 
decreases as $W$ is increased.
At fixed and small enough disorder, 
they both grows with $N$ and seem to approach the standard value $1$ for $N \to \infty$, corresponding to fully delocalized wave-functions. 
Conversely, at fixed and large enough disorder, $D_2$ and $D_1$
decrease to zero as the system size is increased,
implying that $\langle \Upsilon_{2} \rangle , \langle S_\epsilon^{(n)} \rangle \to \textrm{cst}$ for $N \to \infty$, as expected for localized eigenstates.
Although different curves corresponding to different values of $N$ cross much before the localization transition,
a careful analysis of the data shows that the crossing point is in fact slowly but systematically 
drifting towards larger values of $W$ as $N$ is increased.
As for $\langle r \rangle$ and $q^{typ}$, the $n$-dependence of $D_2$ and $D_1$ at fixed $W$ is in fact non-monotonic. 
The characteristic crossover scales over which the non-monotonicity of $D_2$ and $D_1$ is observed is within our numerical accuracy the same 
as the one found above for the level statistics on the scale of the mean level spacing.
This suggests that convergence to the conventional ergodic behavior in the delocalized phase of RRG, with Wigner-Dyson statistics for the energy levels
and $1/N$ scaling of the IPR, is governed by a {\it unique} characteristic correlation volume~\cite{mirlin,levy}.
  
 
\vspace{6pt}

An eigenstate $\ket{m}$ and its coefficients $w_m(i) = |\braket{i}{m}|^2$
can be characterized by the moments (i.e., generalized
IPR)
$\langle \Upsilon_{q} \rangle = \langle \sum_{i=1}^N [w_m(i)]^q \rangle \propto N^{- \tau (q)} \equiv N^{- D_q(q-1)}$.
($\Upsilon_{1} = 1$ for the normalization and
$\langle \Upsilon_{2} \rangle$ is the standard IPR defined above).
For ergodic systems, in the limit $N \to \infty$, all the wave-function amplitudes are
of $O(1/N)$, corresponding
to $\tau(q) = q - 1$. 
Conversely, 
finding that the ratio $D_q = \tau(q)/(q-1)$ depends on $q$ (and is different from one)
is a signatures of non-ergodic states.
In this case, the eigenfunctions
are called {\it multifractal}.
It is customary to characterize the amplitudes $w_m(i)$ by the
spectrum of fractal dimensions $f(\alpha)$,
defined in the following way: The number $\mathcal{N}(\alpha)$
of sites $i$ that have amplitudes scaling as $N^{-\alpha}$
behaves as $\mathcal{N}(\alpha) \simeq N^{f(\alpha)}$.
As a result, one has that:
\begin{displaymath}
\Upsilon_q = \sum_{i=1}^N w_i^q \sim \int \textrm{d} \alpha \exp \left[ \left( f(\alpha) - q \alpha \right) \ln N \right] \, .
\end{displaymath}
Then, in the thermodynamic limit, the saddle point computation of $\Upsilon_q$ leads to the following Legendre transform formula:
\begin{displaymath}
\begin{aligned}
& \alpha = {\rm d} \tau / {\rm d} q \, , \qquad  f^\prime (\alpha) = q \, , \\
& \tau(q) = q \alpha - f(\alpha) \, .
\end{aligned}
\end{displaymath}
$f(\alpha)$ is a convex function of $\alpha$.
The value $q=0$ is associated with the most probable value $\alpha_{m}$
of the wave-function coefficients, where the singularity spectrum
reaches its maximum, $f(\alpha_{m}) = 1$. The value $q=1$ is associated
with the point $\alpha_1$ such that $f(\alpha_1) = \alpha_1$, and $f^\prime(\alpha_1) = 1$.
In the $N \to \infty$ limit, a finite support $0 < \alpha_- < \alpha < \alpha_+$ where
$f(\alpha)>0$ is a signature of multifractality, while for ergodic states,
$f(\alpha) = - \infty$ unless for $\alpha = 1$, where $f(1) = 1$, and
$(\alpha_- < \alpha_1 < 1 < \alpha_{max} < \alpha_+) \to 1$.

The behaviour at low and strong disorder is as expected: At low enough disorder (see App.~\ref{app:multi} and Fig.~\ref{Multi-5}) the support of the singularity spectrum
clearly shrinks as $N$ is increased, and $f_N (\alpha)$ eventually converge to a $\delta$-function for large $N$, 
$\lim_{N \to \infty} f_N (\alpha) = \delta (\alpha - 1)$ (see also Fig.~\ref{Points-5}),
corresponding to full ergodicity; whereas 
in the localized regime (see App.~\ref{app:multi} and Fig.~\ref{Multi-18} for $W=19$), $f_N (\alpha)$ 
gets broader as the system size is increased and
shows a shape which is reminiscent of the triangular form typically observed in the insulating phase.

\begin{figure}
 \includegraphics[angle=0,width=0.48\textwidth]{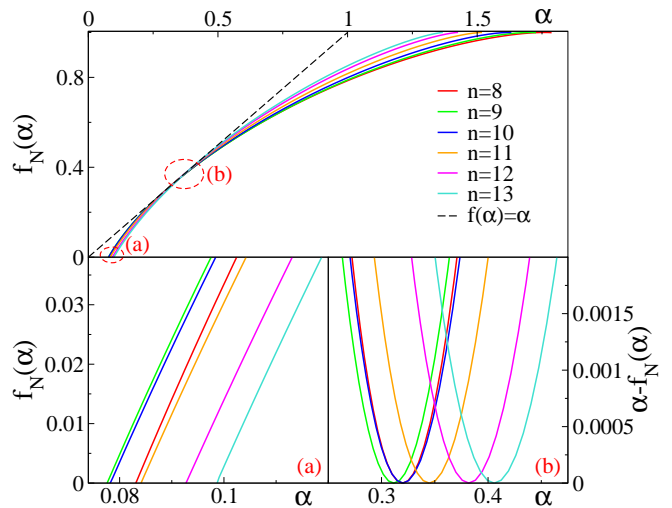}
 \caption{\label{Multi-13} 
 Main panel: Spectrum of fractal dimensions $f_N (\alpha)$ for $W=13$ and for different system sizes 
 $N=2^n$ with $n$ from $8$ to $13$.
 Bottom-left panel: Zoom of the same data in the region (a) close to the lower edge of the support of $f_N (\alpha)$, 
 showing a non-monotonic
 behavior: $\alpha_-$ first moves leftwards for $n<n_c$, and then moves rightwards for $n>n_c$.
 Bottom-right panel: Plot of $\alpha - f_N (\alpha)$ in the region (b), showing the 
 non-monotonicity of $f_N(\alpha)$: $\alpha_1$ first moves leftwards for $n<n_c$ and then moves rightwards 
 for $n>n_c$. }
 \end{figure}

We now focus on the putative intermediate phase. In the top panel of Fig.~\ref{Multi-13} we plot the singularity spectrum for $W=13$---deep in the
crossover non-ergodic regime---and for several system sizes $N=2^n$, with $n$ from $8$ to $13$.
(More information and details 
are given in App.~\ref{app:multi}.)
In the following we will focus in particular on the $N$-dependence of two 
specific points of the singularity spectrum: 
The point $\alpha_1$ (associated to $q=1$) where $f_N (\alpha_1) = \alpha_1$, and $f_N^\prime(\alpha_1) = 1$;
And the lower edge of the support of $f_N(\alpha)$, $\alpha_-$.
The bottom left panel provides a zoom of the same curves in the region (a), close to the lower edge of the support of $f_N (\alpha)$, while
the bottom right panel shows the plots of $\alpha - f_N (\alpha)$ in the region (b), allowing to identify the
position of $\alpha_1$.
These plots clearly demonstrate that the evolution of $f_N (\alpha)$ is non-monotonic:
For small enough sizes (i.e., $n \lesssim n_c (W)$) the support of $f_N(\alpha)$ gets broader, 
and $\alpha_-$ and $\alpha_1$ decrease and as $n$ is increased, as for non-ergodic states.
Conversely, for larger sizes (i.e., $n \gtrsim n_c (W)$) the support of $f_N(\alpha)$ shrinks back,
and $\alpha_-$ and $\alpha_1$ increase with $n$.
A similar behavior is observed in the whole crossover region, $W \gtrsim 10$.
The crossover scale governing the non-monotonic behavior of the singularity spectrum coincides, within our numerical
accuracy, with the one found above from the non-monotonic behavoir of the level statistics and of the IPR.
See Fig.~\ref{min-13} for a summary of the numerical observations discussed above. 
 
\subsection{The characteristic crossover scale} \label{sec:crossover}

The numerical results presented in this section suggest the emergence of a {\it unique} characteristic scale which controls the 
transition from a phase characterized by Poisson statistics--localization--lack-of-ergodicity to one displaying GOE statistics--delocalization--ergodicity for 
the Anderson model on RRGs of finite size.
Such crossover scale is
already very large well below the Anderson localization, resulting in
a broad crossover region where 
finite size effects are extremely important. As mentioned above, in such crossover region  
{\it all} observables and probes introduced in the previous sections 
share the same non-monotonic behavior as a function of the system size~\cite{ioffe1,mirlin,levy}.

 \begin{figure}
 \includegraphics[angle=0,width=0.48\textwidth]{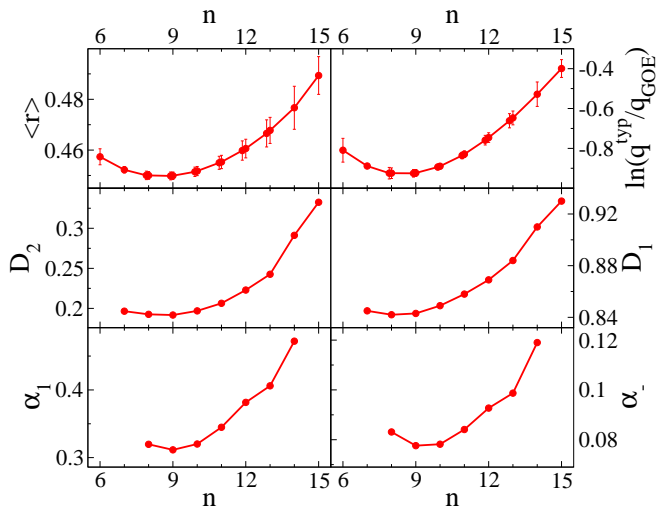}
 \caption{\label{min-13} 
 Non-monotonic behavior of several observables such as $\langle r \rangle$ (top-left panel), $\ln (q^{typ}/q_{\rm GOE})$ 
 (top-right panel),
 $D_2$ (middle-left panel), $D_1$ (middle-right panel), $\alpha_1$ (bottom-left panel), 
 and $\alpha_-$ (bottom-right panel) as a function of $n=\log_2 N$ for $W=13$. The position of the minimum 
 $n_c(W)$ depends weakly on the observable.}
 \end{figure}

This is highlighted in Fig.~\ref{min-13}, where we plot the $n$-dependence of several observables
related, to the statistics of the gaps 
(i.e., 
$\langle r \rangle$ 
and $q^{\rm typ}$), and to wave-functions ergodicity
(i.e., 
$D_2$, 
$D_1$, 
$\alpha_1$, and $\alpha_-$) 
for $W=13$.
All the different curves show a very similar non-monotonic shape. The position of the minimum, $n_c (W)$, seems to depend very weakly on
the choice of the observable.

\begin{figure}
 \includegraphics[angle=0,width=0.46\textwidth]{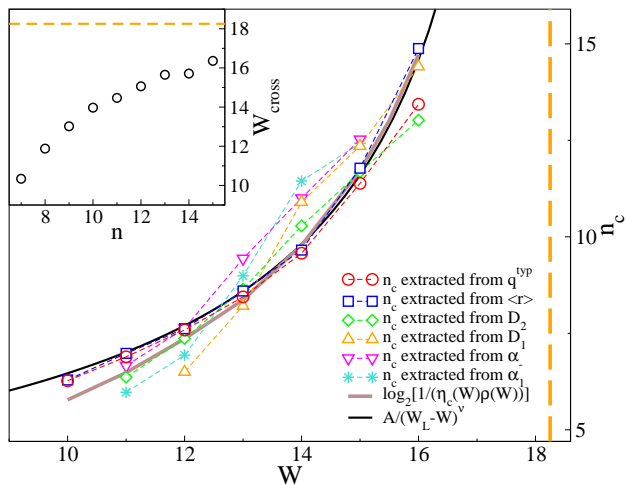}
 \caption{\label{nc} 
 Main panel: Characteristic crossover scales $n_c (W)$ extracted from different
	observables as a function of the disorder $W$. The black curve is a fit of the form $n_c (W) \approx A/(W_L-W)^\nu$
	with $A \approx 20$ and $\nu \approx 0.6$. The gray thick curve represents the estimation of the crossover size
	given by Eq.~(\ref{eq:Nc}).
 Inset: Evolution with $n$ of the crossing point of the curves of $q^{\rm typ}(W)$ of Fig.~\ref{r-q} for two subsequent system sizes.}
 \end{figure}
 
This is confirmed by the main panel of Fig.~\ref{nc}, where we plot the characteristic crossover scales, $n_c (W)$, extracted from the 
different probes, showing that, within our numerical accuracy, they all yield the same dependence on the disorder strength $W$.

The non-monotonic behavior has been interpreted in~\cite{mirlin} in terms of the nature of the Anderson critical point on the RRG, 
which has properties similar to that of the localized phase~\cite{largeD,fyod,efetov,Zirn}, with critical level statistics of Poisson form
and strongly localized critical wave-finctions.
The observables of systems of size $N \ll N_c(W)$ would then first flow upon increasing $N$ towards the critical values, which
tend, for $d \to \infty$, to the ones of the localized phase (i.e., $\langle r \rangle_c = \langle r \rangle_P \approx 0.39$, $q^{\rm typ}_c = 0$, $D_{2,c} = 0$, $D_{1,c} = 0$, $\alpha_{-,c} = 0$, $\alpha_{1,c} = 0$).
Then, when $N$ becomes larger than the correlation volume $N_c$, the observables flow towards their standard values in the delocalized, fully ergodic, phase.

The black curve of Fig.~\ref{nc} shows a fit of the data of the form $n_c \simeq A/(W_L - W)^\nu$, with $A \approx 20$ and $\nu \approx 0.6$,
implying an exponential divergence of the correlation volume at the transition point. 
Although these data are not sufficient to allow for an accurate estimation of $\nu$,  
the value of the exponent is not too far from the one predicted by the supersymmetric analysis, $\nu = 1/2$~\cite{SUSY,fyod}.
Note, however, that recently a different expression has been proposed for $n_c \simeq A/(W_L - W)$, with $\nu = 1$~\cite{ioffe3}.
Our numerical data are clearly too far from $W_L$ to address this controversy.

The gray thick curve of Fig.~\ref{nc} corresponds to the estimation of the crossover scale given by Eq.~(\ref{eq:Nc}) obtained from the convergence of the 
probability distribution of the Local Density of States (LDoS) within the BP approximation 
explained below (see Sec.~\ref{sec:BPLDoS} and Fig.~\ref{fit_length} for more details).

Finally, in the inset of Fig.~\ref{nc} we show the evolution with $n$ of the crossing point of the curves of $q^{\rm typ}(W)$ (presented
in Fig.~\ref{r-q}) for two subsequent system sizes. The crossing point moves very slowly---although in a systematic way---towards
larger values of the disorder as $N$ is increased, and seems to approach $W_L$  in the infinite size limit. 

The numerical results presented here are compatible with the idea that the Anderson model on
the RRG is fully ergodic in the whole delocalized phase in the limit of infinite size, and that standard metallic behavior 
is eventually restored for system sizes larger than the correlation volume, as suggested in~\cite{mirlin,levy} 
and in agreement with the analytical predictions of Refs.~\cite{SUSY,fyod}.
However this conclusion 
is based on the extrapolation of the numerical 
results obtained for 
finite systems, 
and relies on the assumption that no singularity occurs for $N \gg N_c$. 
In fact, this conjecture has been questioned in Refs.~\cite{ioffe1,ioffe2,ioffe3}, where it has been put forward
that there exists a first-order transition in the thermodynamic limit 
between ergodic and non-ergodic states (with a finite jump of, e.g., $D_1$ and $D_2$) at $W_E \approx 10$.
In the following we propose a new approach to deal with this controversy and to answer the open questions 
raised in the introduction. 

\section{BP solution of the iteration equations for the Green's functions on the RRG and on the Cayley tree} 
\label{sec:BP}

As discussed in the introduction, the Anderson model on tree-like structures allows, in principle, for an
exact solution in the limit of infinite lattices~\cite{abou}, 
which yield the probability distribution function of the diagonal elements of the
resolvent matrix, defined as ${\cal G} (z) =  ({\cal H} - z {\cal I}  )^{-1}$.

In order to obtain the recursive equations, the key objects are the so-called {\it cavity} Green's functions, $G_{i \to j} (z)
= [({\cal H}_{i \leftrightarrow j} - z {\cal I}  )^{-1}]_{ii}$, i.e., the
diagonal elements on site $i$ of the resolvent matrix of the modified Hamiltonian ${\cal H}_{i \leftrightarrow j}$ where the edge between
the site $i$ and one of its neighbors $j$ has been removed. 

Take a given site $i$ and its neighbors $\{j_1, \ldots, j_{k+1} \}$ living on an infinite tree. If one removes the site $i$
from the graph, then the sites $\{ j_1, \ldots, j_{k+1} \}$ are uncorrelated, since the lattice would break in $k+1$
semi-infinite disconnected branches. One then obtains (e.g., by Gaussian integration) the following iteration relations for the cavity 
Green's functions~\cite{abou,ourselves}:
\begin{equation} \label{eq:recursion}
G_{i \to j_p}^{-1} (z) = - \epsilon_i - z -
t^2 \!\!\!\! \sum_{j_q \in \partial i / j_p}  \!\!\!\! G_{j_q \to i} (z) \, ,
\end{equation}
where $z=E + i \eta$, $\eta$ is an infinitesimal imaginary regulator which smoothens out the
pole-like singularities in the right hand sides, 
$\epsilon_i$ is the on-site random energy taken from the distribution~(\ref{eq:peps}),
and $\partial i/j$ denotes the set of all $k+1$ neighbors of $i$ except $j$.
(Note that for each site with $k+1$ neighbors one can define $k+1$ cavity Green's functions and $k+1$ recursion relations of this kind.)
After that the solution of Eqs.~(\ref{eq:recursion}) has been found, one can finally obtain the diagonal
elements of the resolvent matrix of the original problem on a given site $i$ as a function of the
cavity Green's functions on the neighboring sites~\cite{ourselves}:
\begin{equation} \label{eq:recursion_final}
{\cal G}_i^{-1} (z) = - \epsilon_i - z - t^2 \!
\sum_{j_q \in \partial i }  \! G_{j_q \to i} (z) \, .
\end{equation}
In the following we will (mostly) focus on the middle of the spectrum ($E=0$) and set $t=1$.

The statistics of the diagonal elements of the resolvent
gives---in the $\eta \to 0^+$ limit---the spectral properties of $\mathcal{H}$.
In particular, the probability distribution of the LDoS at energy $E$ is given by:
\begin{equation} \label{eq:LDoS}
	\begin{aligned}
		\rho_i 
		& = \sum_m | \braket{i}{m} |^2 \, \delta ( E - E_m ) 
		=\lim_{\eta \to 0^+}  \frac{1}{\pi} \, {\rm Im} {\cal G}_i (z) \, ,
	\end{aligned}
\end{equation}
from which the average Density of States (DoS) is simply given by $\rho 
 = (1/N) \sum_i \rho_i 
 = 1/(N \pi) {\rm Tr} \, {\rm Im} {\cal G}$.
Similarly, the IPR can be expressed as:
\begin{equation} \label{eq:IPR}
	\Upsilon_2 
	= \lim_{\eta \to 0^+} \frac{\eta}{\pi \rho(E) N} \sum_{i=1}^N |{\cal G}_{i} (z) |^2 \, .
\end{equation}

Note that Eqs.~(\ref{eq:recursion}) and (\ref{eq:recursion_final}) are {\it exact} on Cayley trees, even for finite lattices
of $n_g$ generations, due to the absence of loops. This is not true instead, on the RRG.
Indeed, in this case, when site $i$ is removed from the graph, the neighbors $\{j_1, \ldots, j_{k+1} \}$ are not truly
decoupled, since they are still connected by some (typically large) loop present somewhere in the system. 
Since the average size of the loops scales as $\ln N$~\cite{wormald}, 
it is reasonable to expect that Eqs.~(\ref{eq:recursion}) and~(\ref{eq:recursion_final}) become 
asymptotically exact in the thermodynamic limit as the cavity Green's functions on sites 
$\{j_1, \ldots, j_{k+1} \}$ become uncorrelated in absence of site $i$ if the typical length of the 
loops which connect them is larger than the correlation length.
This has been in fact proven rigorously in Ref.~\cite{bored} using the local convergence of RRGs to 
Cayley trees.
One can then argue that the recursion equations 
provide an approximate
solution for the diagonal elements of the resolvent matrix for the Anderson model on RRGs of $N$ sites, and that the
quality of the approximation should improve as $N$ is increased. 

Since the Green's functions $G_{i \to j}$ and ${\cal G}_{i}$ are random variables, 
Eqs.~(\ref{eq:recursion}) and (\ref{eq:recursion_final})  
naturally lead to functional equations on their
probability distribution $Q (G)$ and $P({\cal G})$. Let us first focus on the RRG, where the sites of the lattice
are statistically translationally invariant due to the absence of boundaries. From Eq.~(\ref{eq:recursion}) one first 
gets the self-consistent functional equation for the probability distributions of the cavity Green's functions
in the $N \to \infty$ limit (averaged over the on-site disorder and on different realizations of the random lattice):
\begin{equation} \label{eq:PGcav}
\begin{split}
Q (G) &= \! \int \! \textrm{d}p(\epsilon) \prod_{i=1}^k \textrm{d} Q (G_i) \, 
\delta \! \left ( \! G^{-1} \! + \epsilon + z  + \sum_{i=1}^k G_i \! \right) \, ,
\end{split}
\end{equation}
where $p(\epsilon)$ is the probability distribution of the on-site random energy, Eq.~(\ref{eq:peps}).
Once the fixed point of Eq.~(\ref{eq:PGcav}) is obtained, using Eq.~(\ref{eq:recursion_final})
one can compute the probability distribution of the diagonal elements of the resolvent:
\begin{equation} \label{eq:PG}
\begin{split}
	P ({\cal G}) &= \! \int \! \textrm{d}p(\epsilon) \prod_{i=1}^{k+1} \textrm{d} Q (G_i) \, 
	\delta \! \left ( \! {\cal G}^{-1} \! + \epsilon + z + \sum_{i=1}^{k+1} G_i \! \right) \, .
\end{split}
\end{equation}
This set of functional equations can be solved numerically with an arbitrary degree of precision 
using a population dynamics algorithm~\cite{abou,ourselves,ioffe1,ioffe2,ioffe3,PopDyn}.

For Cayley trees, Eqs.~(\ref{eq:PGcav}) and (\ref{eq:PG}) are valid only in the bulk, in the proximity of the root, and at finite $\eta$.
Indeed, due to the presence of the boundary, the lattice is not statistically invariant by translation. 
In order to write the functional iteration equations for the probability distributions of the Green's functions one needs then to take into account 
the position of the sites inside the tree, as explained in detail in App.~\ref{app:cayley}.

In agreement with previous results~\cite{abou,SUSY,fyod,efetov,Zirn,ourselves,aizenmann,semerjian}, we find that in the localized phase, $W > W_L \approx 18.2$, the iteration equations~(\ref{eq:recursion}) and (\ref{eq:PGcav}) are unstable 
with respect to the imaginary regulator $\eta$: $P({\cal G})$ is singular and the average DoS vanishes in the $\eta \to 0^+$ limit. Conversely, in the metallic phase
the probability distribution converges to a stable non-singular distribution function, provided that $\eta < \eta_c(W)$, where $\eta_c(W)$ is an energy scale which 
is finite in the whole delocalized phase and vanishes exponentially as 
$\eta_c \simeq \exp [-{\rm cst}/(W_L-W)^\nu]$ for $W \to W_L$. For $\eta < \eta_c(W)$ the typical value of ${\rm Im} {\cal G}$, 
defied as ${\rm Im} {\cal G}^{\rm typ} = e^{\langle \ln {\rm Im} {\cal G} \rangle}$, also converges
to a $\eta$-independent finite value ${\rm Im} {\cal G}^{\rm typ}_0$ [which vanishes exponentially for $W \to W_L$ with the same exponent $\nu$ 
and behaves as ${\rm Im} {\cal G}^{\rm typ}_0 (W) \propto 
\eta_c (W)^b$, with $b \approx 1.12$ in our numerics, see Fig.~\ref{fit_length}]. 
Similarly, $\langle | {\cal G}_i |^2 \rangle$ converges to
a finite value (which diverges exponentially for $W \to W_L$) in the whole delocalized phase, and the IPR goes to zero.

However, this analysis is carried out when the limit $\eta \to 0^+$ is taken {\it after} the
thermodynamic limit $N \to \infty$. Recently in Refs.~\cite{ioffe1,ioffe2,ioffe3} it has been put forward that taking the limit $N \to \infty$ first
does not allow to detect the existence of delocalized but non-ergodic states (if they exist). Indeed, for multifractal states the wave-functions
typically occupy a fraction of $N^D$ sites (with $0<D<1$), implying the existence of an energy scale $\eta_c (N)$ which decreases as $N^{D-1}$ but 
stays much larger than the mean level spacing $\delta = 1/(\rho N)$. This should be the hallmark of the non-ergodic extended phase~\cite{kravtsov,facoetti,ioffe1,ioffe2,ioffe3}.
In order to deal with this situation, one should instead take the simultaneous limits $N \to \infty$, $\eta \to 0$, with $N^\phi \eta = {\rm cst}$ (with $\phi \le D$).
This motivated the authors of Refs.~\cite{ioffe1,ioffe2,ioffe3} to propose an analytical approximate method, based on RSB and 
called ``inflationary population dynamics'', which consists in modifying the
iteration relations~(\ref{eq:recursion}) and (\ref{eq:PGcav}) in a way that allows to distinguish the multifractal states.

The other issue of taking the $N \to \infty$ limit from the start consists in the fact that several important observables related to the statistics of wave-functions 
and energy levels, are simply not defined on infinite lattices. In order to ascertain their properties one should instead understand their scaling behavior 
with $N$ in the limit of very large sizes.

In this paper we propose a novel and alternative strategy to overcome these issues.
The idea is to solve directly Eqs.~(\ref{eq:recursion})  and~(\ref{eq:recursion_final}) on random instances of large but {\it finite} sizes.
In practice, we first generate the lattice [a random realization of the RRG or the (non-random) Cayley tree] and draw the random on-site energies from Eq.~(\ref{eq:peps}).
Then we find the fixed point of Eqs.~(\ref{eq:recursion}), which becomes a system of $(k+1) N$ coupled equation for the cavity Green's functions~\cite{precisazioneCT}. 
This can
be done iteratively with arbitrary precision in a time which scales linearly with $N$.
Finally, using Eqs.~(\ref{eq:recursion_final}) one obtains the diagonal elements of the resolvent matrix on each site.
We repeat this procedure several times to average over different realizations of the 
disorder.

This strategy is well known in statistical physics and information theory and goes under the name of ``Belief Propagation'' (BP) or ``Message Passing'' algorithm~\cite{mezard}, 
and has been---and still currently is---widely used in particular in the context of random optimization and inference problems,
 and spin glass models on sparse random graphs.
As already said above, the BP approach is exact on the Cayley tree, since in this case Eqs.~(\ref{eq:recursion}) and (\ref{eq:recursion_final}) 
are exact due to the absence of loops.
Conversely, on the RRG the iteration equations 
become 
asymptotically exact in the $N \to \infty$ limit only. 
Although there is a rigorous proof of the convergence of the BP solution for the Anderson model on the RRG in the large $N$ limit~\cite{bored}, there is no rigorous estimate of the error at large but finite $N$.
However, in most cases studied in the literature the BP approach has proven as a powerful, accurate and controlled approximation 
and in general provides 
good estimations of local and average quantities, which improve as the system
size is increased~\cite{mezard}.

 
The BP approach has several advantages:
\begin{itemize}
\item[(a)] The fixed point of Eqs.~(\ref{eq:recursion}) can be found in a time which scales linearly with the size of the system.
This allows to investigate lattices of huge size (e.g., up to $N = 2^{29}$) i.e., several orders of magnitude larger than what can be achieved by the 
most efficient available algorithms of ED~\cite{lemarie}. This allows to overcome finite size effects even deep inside the intermediate supposedly 
		non-ergodic region;
\item[(b)] Although the starting point is provided formally by the same set of local equations both for the RRG and the Cayley tree, the BP algorithm 
	gives in general substantially different fixed point solutions for the two cases, since this it is sensitive to the presence of loops, boundaries, 
	and to the structure of the lattice, thereby allowing to disclose the difference between the two kinds of tree-like graphs;
\item[(c)] Within the BP approach it is natural and straightforward to define observables related to the eigenfunctions and energy level statistics which can be expressed in 
	terms of the Green's functions defined on the sites of a random instance of finite size $N$. 
		Moreover, one can easily investigate the properties of those observables on an energy scale which scales
		in a non-trivial way with the system size (povided that it stays larger than the mean level spacing $\delta$).
		As a consequence, this method allows to unveil the existence of an energy scale which stays larger than $\delta$ but decreases with $N$, which
		is the hallmark of the non-ergodic extended phase~\cite{kravtsov,facoetti,ioffe1,ioffe2,ioffe3}.
\end{itemize}

The rest of the paper is devoted to the BP analysis. In the next section we compare the results found within the BP approach to EDs (up to the accessible system sizes, $N=2^{15}$)   
for 
several values of the disorder, 
and establish its accuracy and the domain of validity. 
In particular, we show that, provided that the imaginary regulator $\eta$ stays larger than the mean level spacing $\delta$, the BP approach yields an excellent
estimation of local and average observables, and also accounts for sample-to-sample fluctuations due to different realizations of the disorder, and spatial
fluctuations due to the local environment.
In Sec.~\ref{sec:BPLDoS} we study the convergence of the probability distribution of the LDoS in the limit of large sizes for the Anderson model on the RRG and
on the Cayley tree.
Finally, in Secs.~\ref{sec:chi} and \ref{sec:K2} we focus on two observables, i.e., the level compressibility~\cite{metha} 
and the overlap correlation function, 
related respectively to the statistics of energy gaps and of wave-functions' 
amplitudes~\cite{kravtsov,metz,mirlin_rev,krav_K2,Alts_chi,chalker_chi,Bogo,chalker_K2,thouless}, 
and analyze their behavior on the RRG and on the Cayley tree.

\subsection{Test of acuracy and domain of applicability of the BP approach} \label{sec:BPtest}

\begin{figure}
 \includegraphics[angle=0,width=0.48\textwidth]{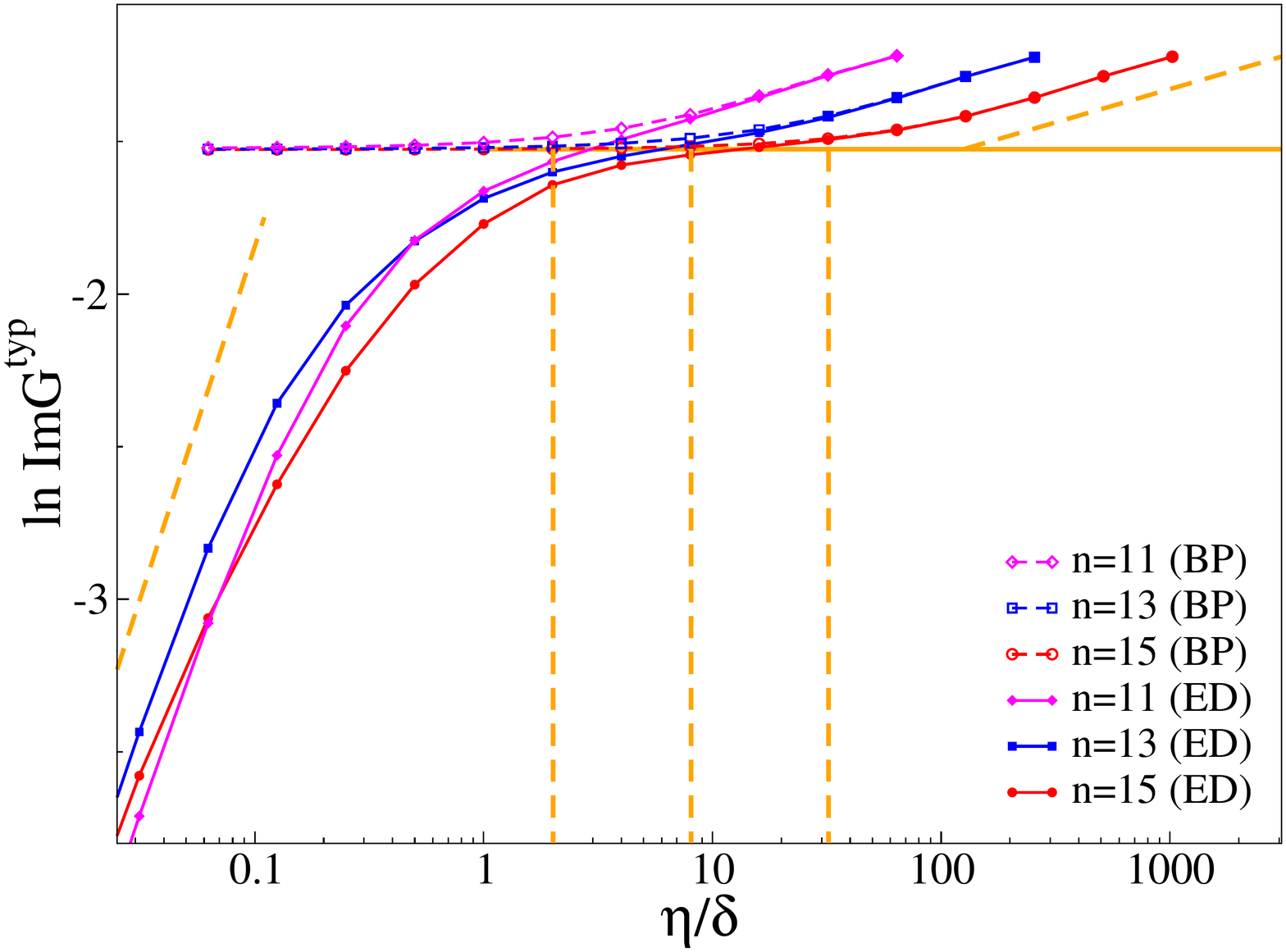}
 \caption{\label{ImGtyp}
	Logarithm of the typical values of the imaginary part of the Green's functions, ${\rm Im} {\cal G}^{\rm typ} = e^{\langle \ln {\rm Im} {\cal G} \rangle}$, as 
	a function of the imaginary regulator measured in units of the mean level spacings $\eta/\delta$, with $\delta = 1/(N \rho)$ for
	three different system sizes, $N=2^n$ with $n=11,13,15$, averaged
	over few realizations of the disorder and of the RRG at $W=6$. The continuous lines and filled symbols
	correspond to the results obtained from ED, whereas the dashed lines and empty symbols correspond to the
	result obtained using the BP approximation on the same sets of random instances. The vertical dashed lines spot the positions of $\eta_c/\delta$ for
	the different system sizes. The full orange dashed lines represent the behavior ${\rm Im} {\cal G}^{\rm typ} \propto \eta$ for $\eta < \delta$
	and ${\rm Im} {\cal G}^{\rm typ} \simeq {\rm Im} {\cal G}^{\rm typ}_0 + (\eta/\eta_c)^\beta$ for $\eta > \delta$. The horizontal full orange line shows
	the $\eta$-independent asymptotic value ${\rm Im} {\cal G}^{\rm typ}_0$
	obtained from Eqs.~(\ref{eq:PGcav}) and (\ref{eq:PG}) for $\eta \ll \eta_c$.}
 \end{figure}

Differently from more standard applications of BP in statistical physics and information theory, in the present 
case the iteration equations~(\ref{eq:recursion}) and 
(\ref{eq:recursion_final}) are ill-defined in the 
limit $\eta \to 0^+$, due to the presence of pole-like singularities in the right hand sides.
One then needs needs to consider the simultaneous limit $N \to \infty$ and $\eta \to 0^+$.
This unusual situation deserves a more careful analysis of the convergence properties of the BP approximation and of its domain of applicability.

In Fig.~\ref{ImGtyp} we plot the behavior of the typical value of the imaginary part of the Green's functions, ${\rm Im} {\cal G}^{\rm typ} = 
\exp[ (1/N) \sum_i \ln {\rm Im} {\cal G}_i ]$ at zero energy as a function of the imaginary regulator $\eta$ measured in units of the mean level spacings
$\delta = 1/(N \rho(W))$ for $N=2^n$ with $n=11,13,15$, averaged over few ($2^{17-n}$) realizations of the on-site disorder and of the RRG, for $W=6$.
The continuous lines and filled symbols show the exact results obtained from the expression of the Greens' functions in terms of the eigenvalues and
eigenvectos, which are obtained from ED:
\[
{\cal G}_i (E + i \eta) = \sum_m | \braket{i}{m} |^2 \frac{E_m - E + i \eta}{(E_m-E)^2 + \eta^2} \, .
\]
One clearly observes three distinct regimes:
\begin{itemize}
	\item[1)] For $\eta < \delta$ the typical LDoS is proportional to the imaginary regulator and vanishes as a constant times $\eta$: If the 
		broadening of the energy levels is smaller than the typical distance between the $\delta$-peaks the system looks as if
		it was localized. In this regime ${\rm Im} {\cal G}^{\rm typ}$ is essentially size-independent, although huge sample-to-sample fluctuations are observed.
	\item[2)] For $\delta < \eta < \eta_c$ the typical value of ${\rm Im} {\cal G}$ reaches a $\eta$-independent (and size-independent provided that $N$ is large enough) plateau value. 
		The threshold $\eta_c(W)$ corresponds to the value of $\eta$ below which the solution of the functional self-consistent equations~(\ref{eq:PGcav}) 
		and (\ref{eq:PG}) for the Green's functions obtained 
                in the thermodynamic limit 
		yields a stable (non-singular) $\eta$-independent function, and the plateau 
		coincides with the value of ${\rm Im} {\cal G}^{\rm typ}$ obtained from this stable
		probability distribution (orange horizontal line, ${\rm Im}G^{\rm typ}_0 \approx -1.525$ for $W=6$). The position of $\eta_c/\delta = N \rho \eta_c$ 
		is highlighted by the vertical dashed lines for the
		different system sizes ($\rho \approx 0.123$ and $\eta_c \approx 8 \cdot 10^{-3}$ for $W=6$). 
		The plateau regime shrinks as the system size is decreased since $\delta$ is proportional to $1/N$. 
		For too small systems (e.g., $N=2^{11}$) the mean level spacing becomes 
		larger than $\eta_c$ and the plateau regime disappears.
	\item[3)] For $\eta > \eta_c$ the typical value of the LDoS grows as ${\rm Im} {\cal G}^{\rm typ} \approx {\rm Im} {\cal G}^{\rm typ}_0 + (\eta/\eta_c)^\beta$.
		The exponent $\beta$ is the same found from Eqs.~(\ref{eq:PGcav}) and (\ref{eq:PG}), and describes the approach of ${\rm Im} {\cal G}^{\rm typ}$ 
		to its limiting value as $\eta$ is decreased below $\eta_c$ ($\beta \approx 0.095$ for $W=6$).
\end{itemize}
Furthermore, we notice that the BP approach (dashed lines and empty symbols) provides a very good approximation 
of the exact result provided that $\eta$ is larger than few mean
level spacings. Conversely, as expected, BP fails completely for $\eta < \delta$.

\begin{figure}
 \includegraphics[angle=0,width=0.46\textwidth]{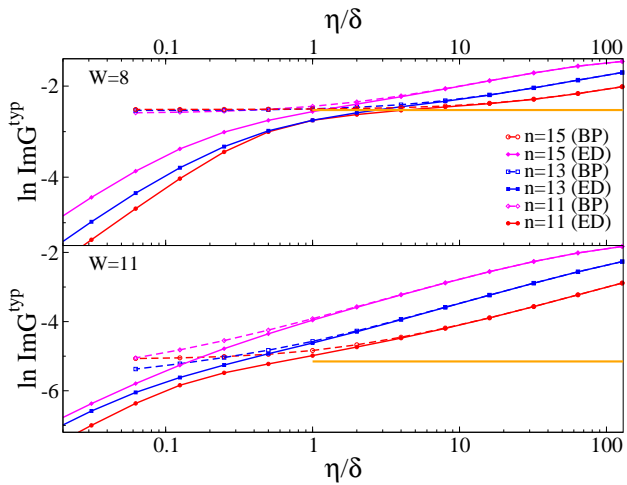}
 \caption{\label{ImGtyp_8_11}
	$\ln {\rm Im} {\cal G}^{\rm typ}$ as a function of $\eta/\delta$ for $W=8$ (top panel) and $W=11$ (bottom panel) for $N=2^n$ with $n=11,13,15$, averaged
        over few realizations of the disorder and of the RRG. Continuous lines and filled symbols
        correspond to the results obtained from ED, whereas the dashed lines and empty symbols correspond to the
        result obtained using the BP approximation on the same sets of random instances.
	The orange horizontal lines correspond to the asymptotic limiting value ${\rm Im} {\cal G}^{\rm typ}_0$
	obtained from Eqs.~(\ref{eq:PGcav}) and (\ref{eq:PG}) for $\eta \ll \eta_c$.}
 \end{figure}

Upon increasing the disorder strength, the average DoS decreases (e.g., $\rho \approx 0.104$ for $W=8$ and $\rho \approx 0.0824$ for $W=11$) and
$\eta_c$ grows extremely fast (e.g., $\eta_c \approx 10^{-3}$ for $W=8$, $\eta_c \approx 8 \cdot 10^{-5}$ for $W=11$). Hence one needs larger
and larger system sizes to be able to observe the plateau. For exemple, at $W=11$ the
plateau regime is not visible even for the largest available system, $N=2^{15}$, while it berely starts to appear at $W=8$ for $N=2^{15}$ (see Fig.~\ref{ImGtyp_8_11}).
In both cases, we still notice an excellent agreement between the exact results and the BP approximation as far as $\eta/ \delta > 1$.

\begin{figure}
 \includegraphics[angle=0,width=0.48\textwidth]{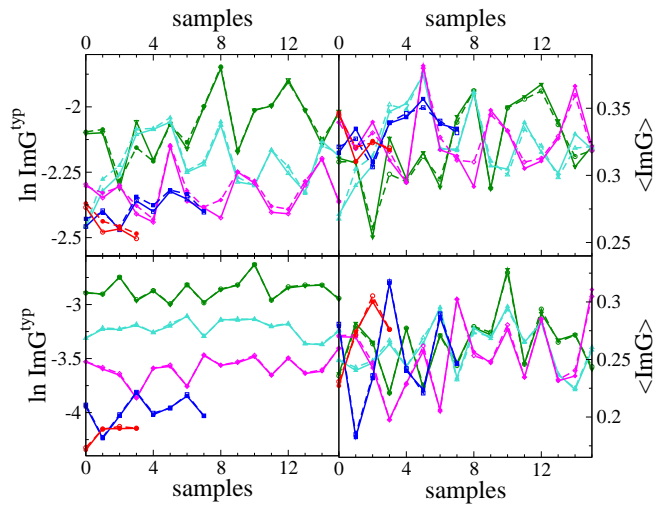}
 \caption{\label{StS}
	$\ln {\rm Im} {\cal G}^{\rm typ}$ (left panels) and $\langle {\rm Im} {\cal G} \rangle$ (right panels) for several random realizations of the Hamiltonian, 
	for $N = 2^n$ with $n=11, \ldots, 15$, and for $W=8$ (top panels) and $W=11$ (bottom panels).
	$\eta$ is set equal to $c \delta$, with $c=8$. Full lines and filled symbols
        correspond to the values obtained from ED, whereas dashed lines and empty symbols represents the
        results obtained using the BP approximation on the same sets of random instances.}
 \end{figure}

On the basis of these observations, from now on we will set the imaginary regulator to be few level spacings, $\eta = c \delta$, with $c \gtrsim 1$.
(recent results~\cite{ioffe_private} suggest that in fact the Anderson model on the RRG might display uncommon
features in the regime $\eta \ll \delta$. Here we do not consider such regime and focus on the more 
standard situation $\eta > \delta$ only.)

\begin{figure}
 \includegraphics[angle=0,width=0.44\textwidth]{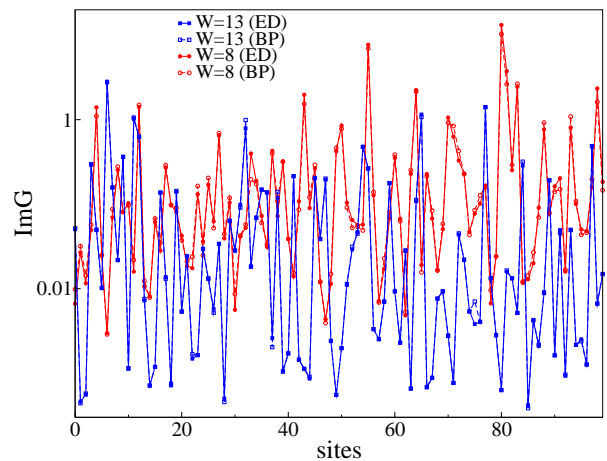}
 \caption{\label{ImGlocal}
	${\rm Im} {\cal G}_i$ for the first $100$ sites $i$ of a RRG of $2^{14}$ sites for $W=8$ (red) and $W=13$ (blue) and for $\eta = c \delta$ with $c=8$.
	Full lines and filled symbols  correspond to the values obtained from ED, whereas dashed lines and empty symbols represents the
	results obtained using the BP approximation.}
 \end{figure}

In Fig.~\ref{StS} we show the values of the typical (left panels) and average (right panels) DoS for several random realizations of the Hamiltonian~(\ref{eq:H}), 
for two values of the
disorder, $W=8$ (top panels) and $W=11$ (bottom panels), for five different system sizes, $N = 2^n$ with $n=11, \ldots, 15$, and $c=8$.
The contiuous lines and filled symbols corresponds to the values obtained from ED, while the dashed lines and open symbols represent the results found with the
BP approximation. These data show that BP correctly reproduces not only average quantities but also accounts for sample-to-sample fluctuations
in an extremely satisfactory way. Moreover, one can check that the relative error of the BP results on average quantities decreases with $N$ (roughly as $1/\sqrt{N}$).
We also find that the relative error of the BP approximation decreases with the disorder strength (see also Fig.~\ref{ImGlocal}). 
Although this might seem surprising at first, 
one can 
rationalize this obervation by recalling that the errors done by the BP approximation are due to the presence of 
loops of finite size (i.e., smaller than the collrelation length $\ln N_c$)
where a resonance between two sites belonging to the same loop occurs~\cite{ioffe_private}.
The number of such loops 
in the large $N$ limit is given asymptotically by some known distribution function and stay of $O(1)$~\cite{wormald}.
When $W$ is increased, the propability that two sites belonging to a short loop are in resonance decreases, 
and the accuracy of the BP results improves.

In Fig.~\ref{ImGlocal} we plot ${\rm Im} {\cal G}_i$ for the first $100$ sites $i$ of a specific realization of a RRG of $2^{14}$ sites and of
the on-site disorder for $W=8$ and $W=13$ (and $c=8$ as before). We compare again the values obtained from ED with the results of the 
BP approximation, showing that BP provides an excellent estimations also of the {\it local} Green's functions, and is able to describe the spatial fluctuations
due to the local environment.
Only very small discrepancies on some specific sites are observed. Those sites are likely to belong to short loops and to be in resonance with
another site of the same loops.

\begin{figure}
 \includegraphics[angle=0,width=0.48\textwidth]{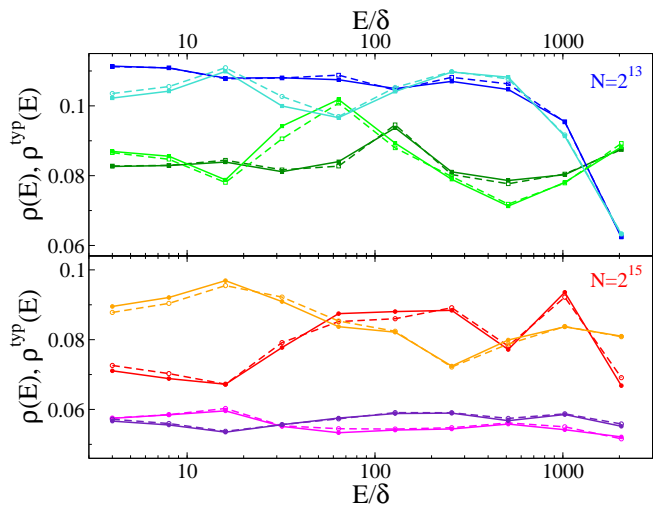}
 \caption{\label{rhoE}
	Average (red, orange, light green and dark green) and typical (violet, indigo, bule, turquoise) DoS, $\rho (E) = \langle {\rm Im} {\cal G} \rangle / \pi$ and 
	$\rho^{\rm typ} (E) = e^{\langle \ln {\rm Im} {\cal G} \rangle} / \langle {\rm Im} {\cal G} \rangle$, as a function of $E / \delta$ 
	at $W=11$ for two specific realizations of the RRG and
	of the random energies of $N=2^{13}$ (top panel) and $N=2^{15}$ (bottom panel) sites, and for $\eta = c \delta$ with $c=8$.
        Full lines and filled symbols  correspond to the values obtained from ED, whereas dashed lines and empty symbols represents the
        results obtained using the BP approximation.}
 \end{figure}

Finally, in Fig.~\ref{rhoE} we plot the average DoS $\rho (E) = \langle {\rm Im} {\cal G} (E) \rangle / \pi$ and the typical DoS
$\rho^{\rm typ} (E) = e^{\langle \ln {\rm Im} {\cal G} (E) \rangle} / \langle {\rm Im} {\cal G} (E) \rangle$ as a function of the energy $E$ measured
in units of the mean level spacings $\delta$, for $W=11$ and for two different realizations of the on-site disorder and of the RRG of $N=2^{13}$ (top panel)
and $N=2^{15}$ (bottom panel) sites (and for $c=8$). Once again, the comparison between the BP approximations with the values obtained from ED is very good,
showing that BP reproduces correctly the fluctuations of the DoS over all range of energies, from the order of the mean level spacing up to energies of the order of the band-width,
and that the quality of the approximation improves as the system size is increased.

All in all, these findings shows that the BP approach yields a powerful, efficient and accurate approximation for the Green's functions
of the Anderson model on the RRG, not only at the level of average quantities, but also at the local scale,
provided that the imaginary regulator is scales as the mean level spacing times a constant of order $1$.
It also reproduces correctly the fluctuations between different random instances due to different random realization of the graph and of the quenced diagonal
elements of the Hamiltonian, and works nicely over the whole energy range from energies of order $1/N$ up to energies of order $1$.
The relative error of the BP approximation decreases as the system size $N$ and the disorder strength $W$ are increased.

\subsection{Convergence of the distribution of the LDoS} \label{sec:BPLDoS}

\begin{figure}
 \includegraphics[angle=0,width=0.46\textwidth]{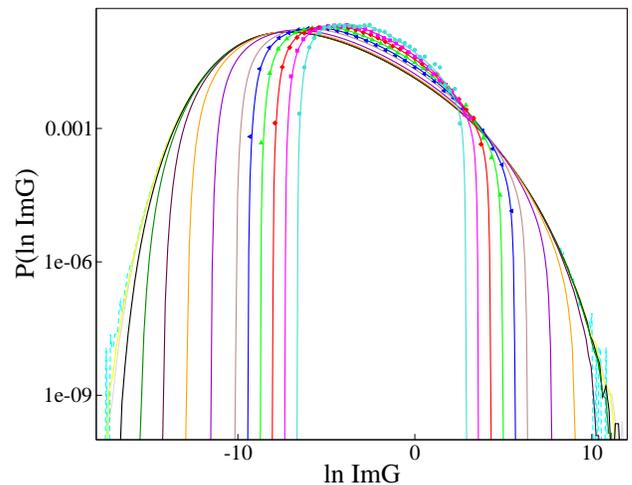}
 \caption{\label{PImGRRG}
	Probability distribution functions of $\ln {\rm Im} {\cal G}$ at $W=12$ for the Anderson model on RRGs  of several sizes $N=2^{n}$, with $n=11$ (turquoise), 
	$n=12$ (magenta), $n=13$ (red), $n=14$ (light green), $n=15$ (blue), $n=16$ (brown), $n=18$ (violet), $n=20$ (orange), 
	$n=22$ (maroon), $n=24$ (black), $n=26$ (dark green), $n=28$ (black), and $n=29$ (yellow), 
	averaged over several realizations of the RRG and of the on-site disorder (and for $c=8$).
	Full curves correspond to the results of the BP approach, whereas symbols represent the PDFs obtained from ED (but averaged 
	over $2^{17-n}$ realizations only).
	The dashed light blue curve shows the solution of the functional self-consistent equations valid in the thermodynamic 
	limit, Eqs.~(\ref{eq:PGcav}) and (\ref{eq:PG}), found via
	population dynamics for $\eta < \eta_c$.}
 \end{figure}

\begin{figure}
 \includegraphics[angle=0,width=0.48\textwidth]{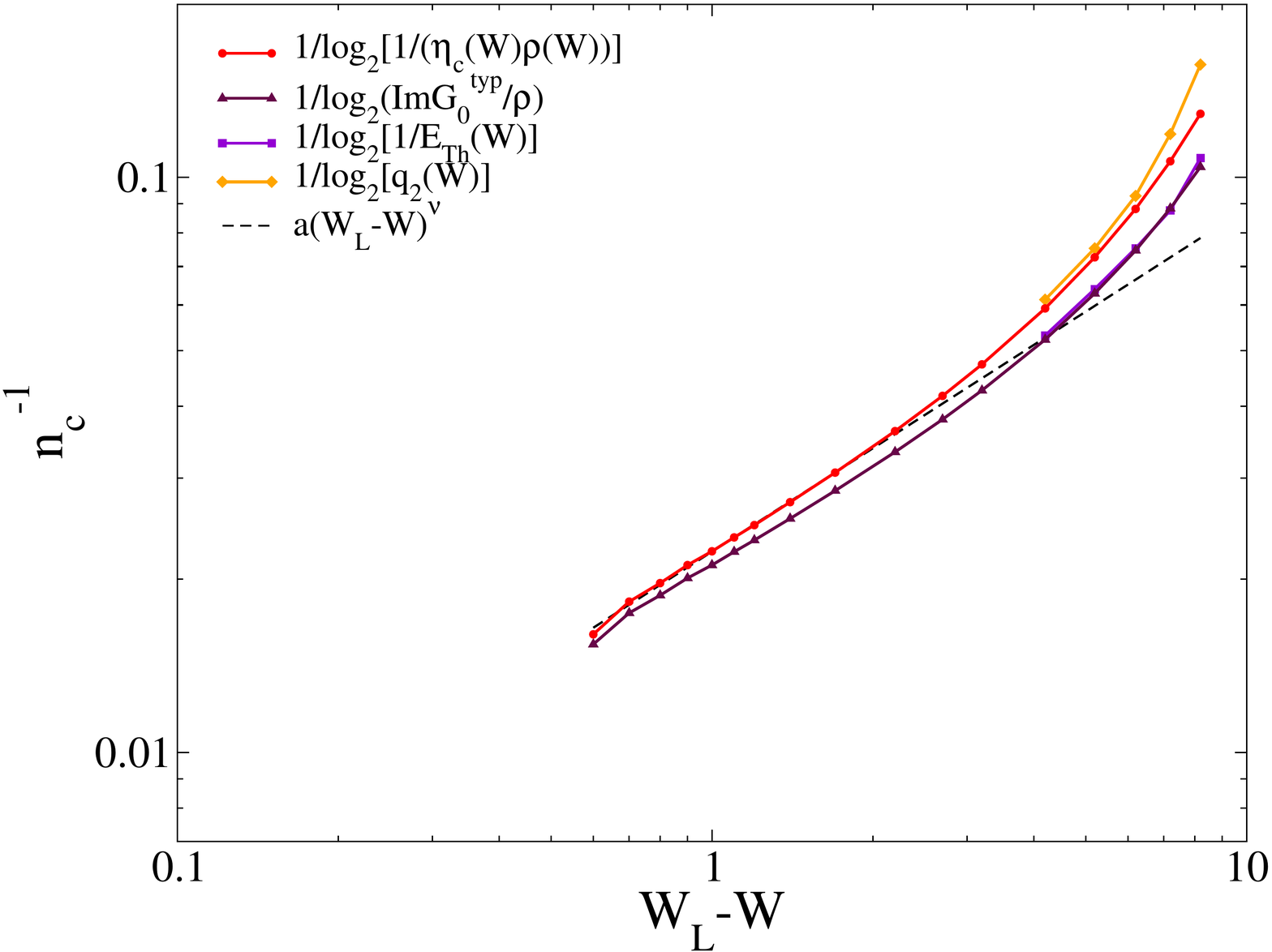}
 \caption{\label{fit_length} 
 Inverse of the characteristic crossover length $n_c^{-1} = 1/\log_2 N_c(W)$ (red curve) obtained from the convergence of the distribution of the LDoS, Eq.~(\ref{eq:Nc}). 
 The dashed black line corresponds to a fit of the data as $n_c^{-1} \approx a(W_L-W)^\nu$, with $a \approx 0.022$, $\nu \approx 0.6$, and $W_L \simeq 18.2$.
 The plot also shows the inverse of the logarithm of the asymptotic value of ${\rm Im}{\cal G}^{\rm typ}$, $1/\log_2 ({\rm Im}{\cal G}^{\rm typ}_0/\rho)$ (maroon), of the inverse Thouless energy, $1/\log_2 E_{Th}^{-1}$ (magenta), and of the value of the plateau if the overlap correlation function $K_2 (E)$ at small energy, $1/\log_2 q_2$ (violet), see Sec.~\ref{sec:K2} for a precise definition of the last two quantities.}
 \end{figure}

In this section we focus on the convergence of the probability distribution of the LDoS obtained from the BP approach for large but {\it finite} systems.
As mentioned above, one of the advantage of BP is that the system of coupled equations~(\ref{eq:recursion})  and~(\ref{eq:recursion_final})
can be easily solved by iteration in a linear time in $N$, thereby allowing to access system size several order of magnitude larger than
the ones currently accessible via ED.
In Fig.~\ref{PImGRRG} we show the probability distributions of the imaginary part of the Green's functions of the
Anderson model on the RRG, for $N=2^n$ with $n = 11, \ldots, 29$ at $W=12$ (deep into the putative
delocalized non-ergodic phase), 
averaged over many independent realization of the disorder (as in the previous section we set $\eta = c \delta$ with $c=8$).
We observe that:
\begin{itemize}
	\item $P(\ln {\rm Im} {\cal G})$ converges to a stable, non-singular, size independent distribution for large enough sizes (say, for $N \gtrsim 2^{28}$);
	\item Convergence occurs when $\eta$ becomes smaller than an energy scale $\eta_c (W)$ which remains finite, and 
		which coincides with the scale below
		which the solution of Eqs.~(\ref{eq:PGcav}) and (\ref{eq:PG}) becomes stationary;
	\item The stationary probability distribution at large $N$
		turns out to be the same (within our numerical accuracy) as the one found from Eqs.~(\ref{eq:PGcav}) and (\ref{eq:PG}) for $\eta < \eta_c$;
	\item For the system sizes accessible via ED ($N=2^{11}, \ldots, 2^{15}$) we find an excellent agreement between the BP results
		and the exact distributions;
	\item Since $(1/N) \sum_{i=1}^N | {\cal G}_i |^2$ converges to a size-independent finite value, from Eq.~(\ref{eq:IPR}) one has that the
	IPR 
	goes to zero as $\eta \propto 1/N$. 
	\item We find the very same scenario for all values of the disored strength $W \lesssim 13.5$. For larger values of the disorder the
		correlation volume $N_c (W)$ which would be required to observe the convergence to a stationary distribution, 
		\begin{equation} \label{eq:Nc}
		N_{c} (W) = \frac{c}{\rho(W) \eta_c(W)} \, ,
		\end{equation} 
		becomes exceedingly large, due to the fact that $\eta_c$ becomes exponentially small as one move closer to $W_L$.
		Interestingly, such estimation of the crossover size obtained from the convergence of the probability distribution of the LDoS within the BP
		approach, Eq.~(\ref{eq:Nc}), is plotted in Fig.~\ref{nc} as a gray thick line, showing that it accounts quite well for the scale
		on which the ED data exhibit the non monotonicity. In Fig.~\ref{fit_length} we plot the inverse of the characteristic crossover length controlling the convergence of the LDoS, $n_c^{-1} = 1/\log_2 N_c(W)$, given by Eq.~(\ref{eq:Nc}), as a function of the distance from the Anderson localization $W_L-W$, together with the inverse of the logarithm of the asymptotic value of ${\rm Im}{\cal G}^{\rm typ}$ found for $\eta < \eta_c$, $1/\log_2 ({\rm Im}{\cal G}^{\rm typ}_0/\rho)$. 
		The figure also shows the behavior of the inverse of the logarithm of
		the inverse of the Thouless energy, $1/\log_2 E_{Th}^{-1}$, and of the inverse of the logarithm of the plateau at small energies of the function $K_2 (E)$, $1/\log_2 q_2$ (see Sec.~\ref{sec:K2} for a precise definition of these quantities). Within the BP approximation we find that $N_c \propto {\rm Im}{\cal G}^{\rm typ}_0 \propto E_{Th}^{-1} \propto q_2$, implying that the convergence of the spectral statistics is dominated by a {\it unique} characteristic volume which diverges exponentially fast as $W_L$ is approached.
\end{itemize}

\begin{figure}
 \includegraphics[angle=0,width=0.46\textwidth]{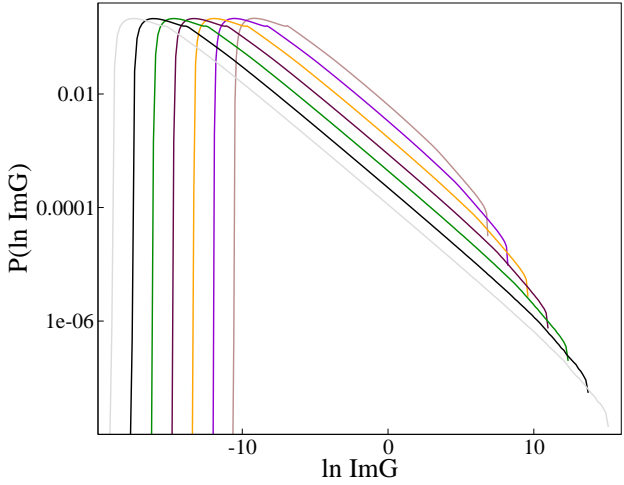}
 \caption{\label{PImGCAY}
        Probability distribution functions of $\ln {\rm Im} {\cal G}$ at $W=12$ for the Anderson mdodel on the loop-less Cayley tree 
	of $n_g$ generations, with $n_g=15$ (brown), $n_g=17$ (violet), $n_g=19$ (orange),
        $n_g=21$ (maroon), $n_g=23$ (dark green), $n_g=25$ (black), and $n_g=27$ (gray), 
	averaged over several realizations of the on-site disorder (and for $c=8$). 
	The typical value of ${\rm Im} {\cal G}$ decreases as $N^{1-D_1}$ (with $D_1 \approx 0.98$ for $W=12$). The tails of the distributions exhibit a power-law
	$P({\rm Im} {\cal G}) \simeq {\rm cst}/({\rm Im} {\cal G})^{1 + \mu}$ with an exponent $\mu \approx 0.52$ up to the cut-off at $1/\eta$.}
 \end{figure}

In Fig.~\ref{PImGCAY} we show the probability distributions of the imaginary part of the Green's functions of the
Anderson model on finite loop-less Cayley trees of $n_g$ generations at the same value of the disorder, $W=12$, showing that 
the situation is drastically different in this case. $P(\ln {\rm Im} {\cal G})$ never converges to a stable distribution, and keeps evolving
as $N$ is increased.
The typical value of ${\rm Im} {\cal G}$ decreases as $N^{1-D_1}$ (or, equivalently, as $\eta^{D_1-1}$) with a non-trivial disorder-dependent
spectral fractal dimension $D_1$ between $0$ and $1$ (see below). 
The average value of ${\rm Im} {\cal G}$ instead approaches a $N$-independent value (corresponding to $\pi$ times
the average DoS), due to the presence of fat tails at large values of ${\rm Im} {\cal G}$: $P({\rm Im} {\cal G}) \simeq {\rm cst}/({\rm Im} {\cal G})^{1 + \mu}$,
with $1/2 \le \mu \le 1$. 
These are precisely the distinctive features which characterize the non-ergodic extended phase and the multifractal states.

\subsection{Spectral fractal exponents} \label{sec:BPD}

\begin{figure}
 \includegraphics[angle=0,width=0.48\textwidth]{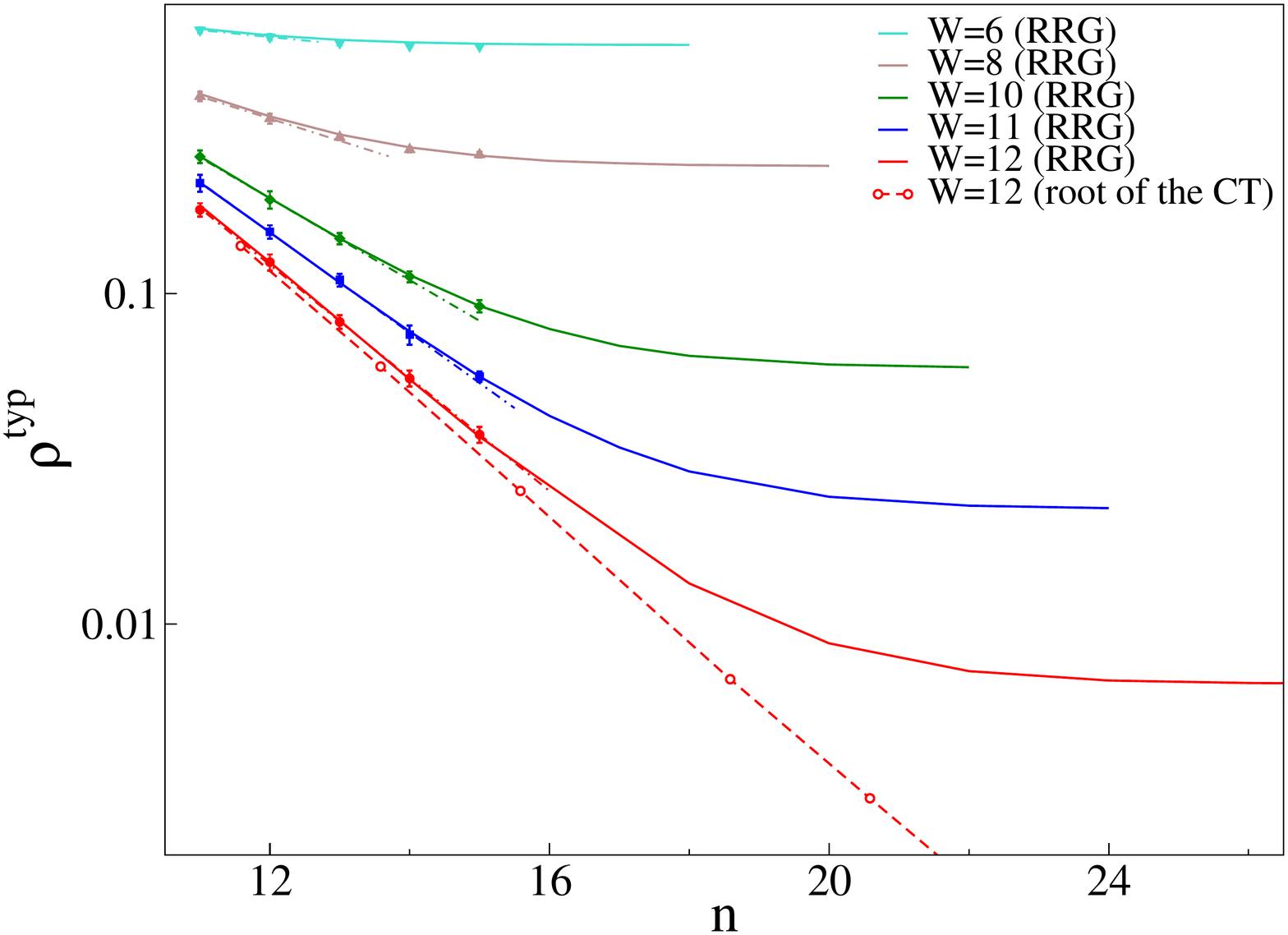}
 \caption{\label{rho_typ_RRG}
	Typical value of the DoS, averaged over many independent realizations of the on-site disorder and of the RRG, 
	as a function of the system size $n = \ln N / \ln 2$ for several values of the disorder strength.
	The continuous curves give the results of the BP approximation. The symbols correspond to the values obtained
	from ED up to the largest accessible system sizes ($N=2^{15}$). The dotted-dashed black line shows the fits $\rho^{\rm typ} \propto N^{1-D_1}$
	over the range of $N$ where one observes an apparent power-law dependence and a multifractal behavior.
	The red dashed straight line and empty circles represents the behavior of $\rho^{\rm typ}$ as a function of $n=\ln N / \ln 2$ 
	for the Anderson model on the Cayley tree at $W=12$ and measured at the root of the tree.}
 \end{figure}

\begin{figure}
 \includegraphics[angle=0,width=0.48\textwidth]{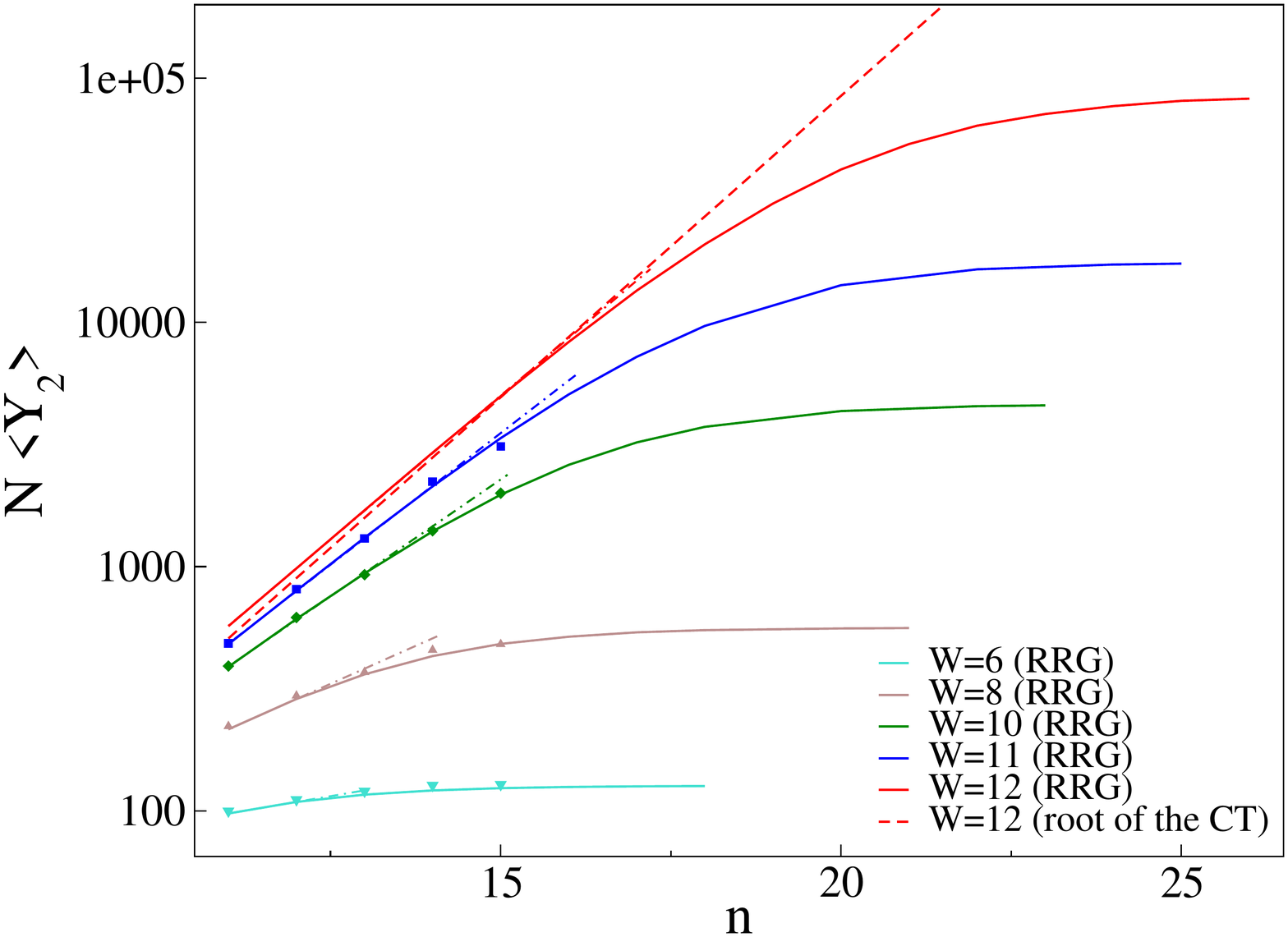}
 \caption{\label{IPR_RRG}
        $N$ times the IPR, averaged over many independent realizations of the on-site disorder and of the RRG,
        as a function of the system size $n = \ln N / \ln 2$ for several values of the disorder strength.
        The continuous curves give the results of the BP approximation. The symbols correspond to the values obtained
        from ED up to the largest accessible system sizes ($N=2^{15}$). The dotted-dashed black line shows the fits $\langle 
	\Upsilon_2 \rangle \propto N^{-D_2}$
        over the range of $N$ where one observes an apparent power-law dependence and a multifractal behavior.
	The red dashed straight line represents the behavior of 
	$N \langle \Upsilon_2 \rangle$ as a function of $n=\ln N / \ln 2$
        for the Anderson model on the Cayley tree at $W=12$ and measured at the root of the tree.}
 \end{figure}

\begin{figure}
 \includegraphics[angle=0,width=0.48\textwidth]{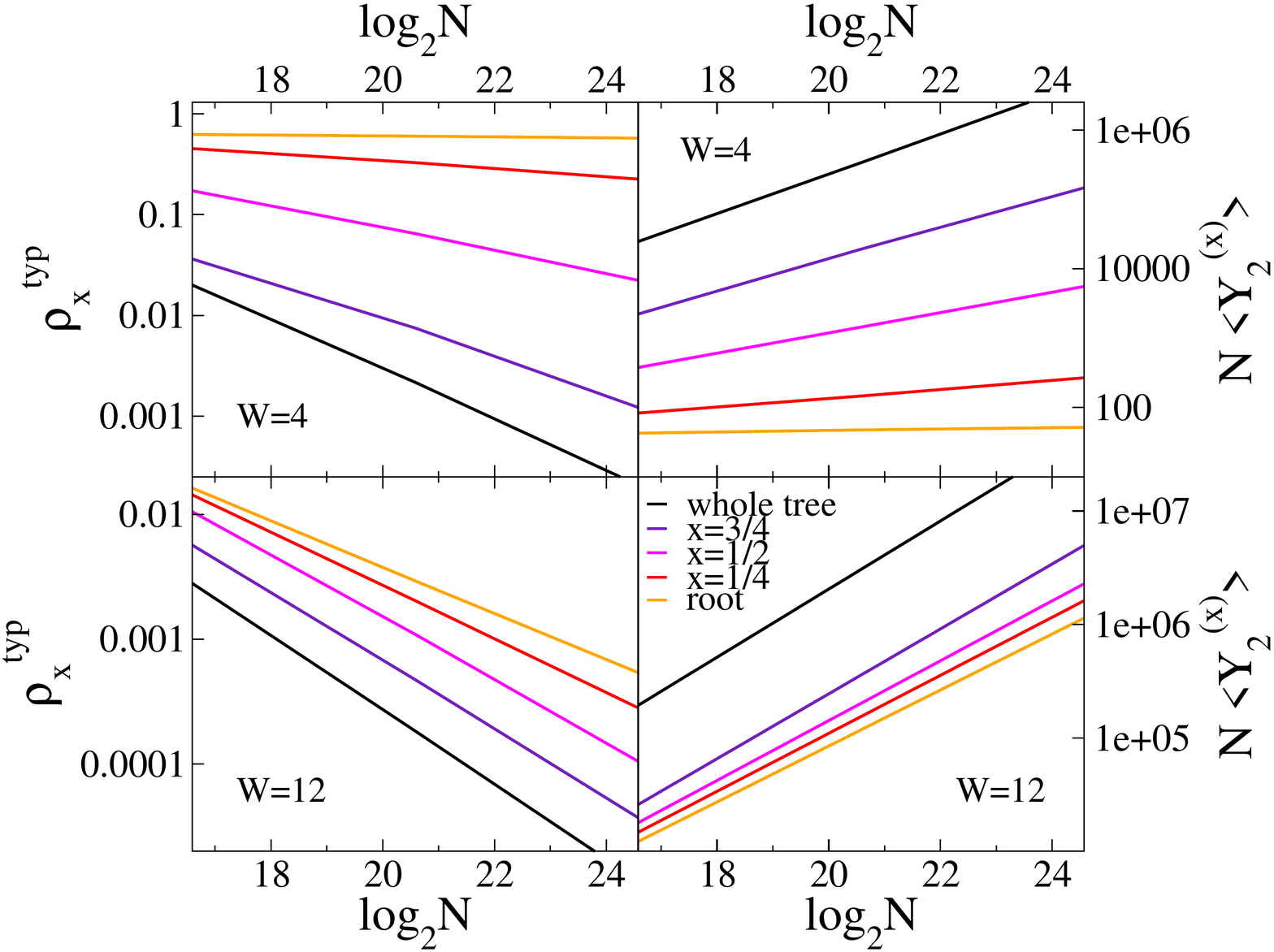}
 \caption{\label{IPR_CT}
 Typical value of the DoS at position $x$, $\rho^{\rm typ}_x$ (left panels), and $N$ times the IPR at position $x$, $N \langle \Upsilon_2^{(x)} \rangle$ (right panels), averaged over many independent realizations of the on-site disorder, 
	as a function of the system size $n = \ln N / \ln 2$ for $W=4$ (top panels) and $W=12$ (bottom panels) for the Anderson model on the Cayley tree     
        at the root of the tree, $x=0$ (orange), for $x=1/4$ (red), $x=1/2$ (magenta), $x=3/4$ (violet), and for the whole tree (black) respectively.}
 \end{figure}

The drastically different behavior observed on the RRG and on the Cayley tree is clearly illustrated by Figs.~\ref{rho_typ_RRG},~\ref{IPR_RRG}, and~\ref{IPR_CT}.
In Figs.~\ref{rho_typ_RRG} and~\ref{IPR_RRG}
we show the evolution with the system size of the typical DoS, $\rho^{\rm typ} = e^{\langle \ln {\rm Im} {\cal G} \rangle} 
/ \langle {\rm Im} {\cal G} \rangle$ and
of the IPR [Eq.~(\ref{eq:IPR})], averaged over several realizations of the disorder and of the RRG for several values of $W$,
which give access directly to the fractal exponents $D_1$~\cite{ioffe1,ioffe2,ioffe3} and $D_2$.

The plots show that for small enough system sizes 
the Anderson model on the RRG behaves {\it as if} it was in a non-ergodic extended phase: 
$\rho^{\rm typ}$ and $\langle \Upsilon_2 \rangle$ show apparent power-law behaviors, $\rho^{\rm typ} \propto N^{1-D_1}$
and $\langle \Upsilon_2 \rangle \propto N^{-D_2}$. 
However, for large enough sizes [i.e., larger than
the crossover scale $N_c (W) = c/(\rho(W) \eta_c(W))$, Eq.~(\ref{eq:Nc})] the $N$-dependence of $\rho^{\rm typ}$  
and $\langle \Upsilon_2 \rangle$ saturates to a $N$-independent value---which coincides with the ones found from the solution of 
Eqs.~(\ref{eq:PGcav}) and~(\ref{eq:PG})---and ergodicity is restored.
Again we observe an excellent agreement between the BP approximation (continuous curves) 
and the results obtained from ED (filled simbols) up to the accessible system sizes.
Yet, due to the fact that the crossover volume $N_c(W)$ grows exponentially fast as $W$ is increased and is already very large far below $W_L$,
the recovery of ergodicity is only visible via ED for moderately weak disorder, $W \lesssim 8$.

It is important to stress that the properties of the corossover region are highly unusual, as
the apparently non-ergodic behavior can be characterized by a set of multifractal exponents, e.g., $D_1$ and $D_2$, 
which are well-defined 
over a broad range of $N$ and depend on the disorder in a non-trivial way. In order to interpret these results, 
we also plot the evolution with the system size of the typical DoS and of the IPR at $W=12$ at the root of Cayley trees of $n_g$ generations (see below for a precise definition of these quantities), 
showing that the spectral fractal dimensions found at the
root of the Cayley tree turn out to be suprisingly close to the apparent multifractal exponents observed on the RRG for $N<N_c$.
The same behavior is found at all disorder strengths.

In fact, as discussed above, the Cayley tree is not translationally invariant and
sites at different distances from the root are not equivalent, it is instructive to study 
the behavior of the typical DoS and of the
IPR at a given depth $\ell$:
\[
	\begin{aligned}
		\rho^{\rm typ}_\ell & = \frac{e^{\langle \ln {\rm Im} {\cal G}_\ell \rangle}}
		{\langle {\rm Im} {\cal G} \rangle} \, , \\ 
		\Upsilon_2^{(\ell)} & = \lim_{\eta \to 0^+} \frac{\eta}{\pi \rho N_\ell} \sum_{i_\ell = 1}^{N_\ell} | {\cal G}_{i_\ell} |^2 \, ,
	\end{aligned}
\]
where $N_\ell = (k+1)k^{\ell-1}$ is the total number of 
sites $i_\ell$ belonging to the $\ell$-the generation of the tree.
As already noticed in~\cite{mirlin_cayley} the appropriate scaling variable characterizing the position of the sites on
a Cayley tree of $n_g$ generations is the dimensionless distance from the root, $x = \ell/n_g$, with $0 \le x \le 1$. 
It was shown in~\cite{mirlin_cayley} that for a given disorder strength $x$ controls the spectrum of wave-functions' multifractal exponents.
In Fig.~\ref{IPR_CT} we plot the evolution with the system size of $\rho^{\rm typ}_x$ and
$N \langle \Upsilon_2^{(x)} \rangle$ at the root of the tree, $x=0$ (orange), for $x=1/4$ (red), $x=1/2$ (magenta), $x=3/4$ (violet), and 
for the whole tree (black) at $W=4$ and $W=12$
showing that the Anderson model on the Cayley tree displays a non-ergodic multifractal behavior at all scales in the whole
delocalized phase (except at small enough disorder and sifficiently close to the root~\cite{mirlin_cayley,DPRM_CT}), $\rho^{\rm typ}_x
\propto N_x^{1-D_1^{(x)}}$ and $\langle \Upsilon_2^{(x)} \rangle \propto N_x^{-D_2^{(x)}}$,
with spectral fractal dimensions $D_1^{(x)} (W)$ and $D_2^{(x)} (W)$ which decrease as $x$ is increased 
(i.e., when one moves closer to the boundary of the tree,
consistently with localization of wave-functions at the boundary~\cite{canopy,mirlin_cayley})
and as $W$ is increased (the spectral fractal dimensions $D_{1,2}^{(x)}$ all vanish at the Anderson transition at $W_L$).

\begin{figure}
 \includegraphics[angle=0,width=0.45\textwidth]{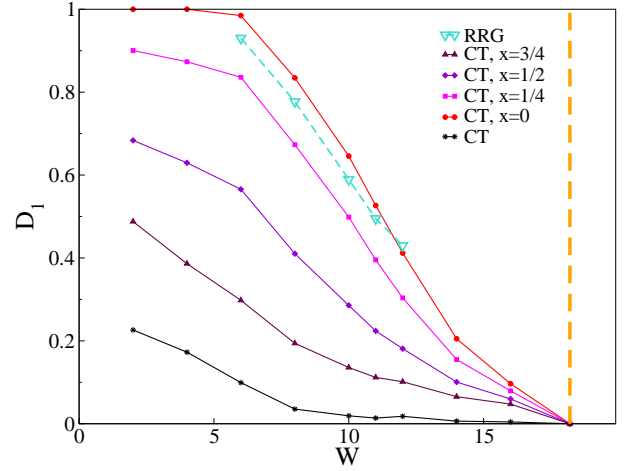}
 \caption{\label{Dmulti}
	Spectral fractal dimensions $D_1$ (black lines and stars) and $D_1^{(x)}$ as a function of $W$ for $x=0$ 
	(i.e., at the root, red line and circles), 
	$x=1/4$ (magenta line and squares), $x=1/2$ (violet line and diamonds), and $x=3/4$ (maroon line and up triangles)
	for the Anderson model on the Cayley tree. The turquoise dashed line and empty down triangles show the values of $D_1$
	on the RRG, measured in the non-ergodic crossover region, i.e., $N < N_c(W)$.
	The orange dashed vertical line represents the location of the Anderson transition, where all spectral fractal dimensions vanish.}
 \end{figure}

\begin{figure}
 \includegraphics[angle=0,width=0.45\textwidth]{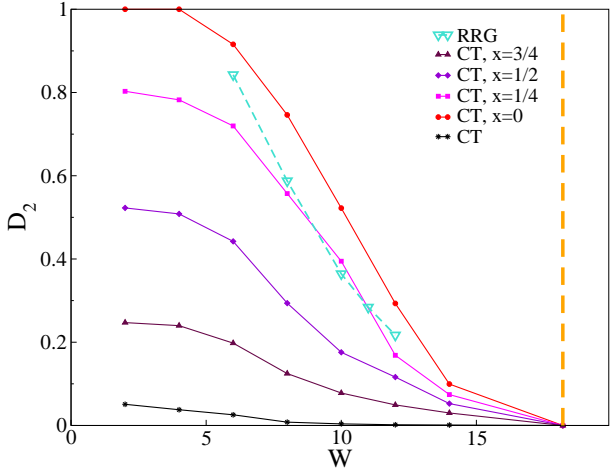}
 \caption{\label{D2multi}
	Spectral fractal dimensions $D_2$ (black lines and stars) and $D_2^{(x)}$ as a function of $W$ for $x=0$ (i.e., at the root, red line and circles), 
        $x=1/4$ (magenta line and squares), $x=1/2$ (violet line and diamonds), and $x=3/4$ (maroon line and up triangles)
        for the Anderson model on the Cayley tree. The turquoise dashed line and empty down triangles show the values of $D_2$
        on the RRG, measured in the non-ergodic crossover region, i.e., $N < N_c(W)$.
        The orange dashed vertical line represents the location of the Anderson transition, where all spectral fractal dimensions vanish.}
 \end{figure}

In Figs.~\ref{Dmulti} and~\ref{D2multi} we plot the behavior of $D_1$ and $D_1^{(x)} (W)$ as a function of the disorder strength
for the Anderson model on the Cayley tree  
for four different positions inside the lattice, $x=0$, $x=1/4$, $x=1/2$, and $x=3/4$.
The spectral fractal dimensions $D_1$ and $D_2$ of the whole tree are controlled by the one of the leaves 
($D_{1,2}^{(x=1)}$), 
since the boundary contains roughly half of the total sites.
We also show on the same plot the apparent spectral fractal dimensions $D_1$ and $D_2$ measured on the RRG in the non-ergodic
crossover region, for $N < N_c (W)$, which, as anticipated above, turn out to be close to the spectral fractal dimension found at the
root of the Cayley tree, $D_{1,2} (W) \simeq D_{1,2}^{(x=0)} (W)$, at the same disorder strength.
(Note that the root of the Cayley tree displays a transition at $W=W_T \approx 6$, below which we find that $D_{1,2}^{(x=0)} = 1$, see, e.g., the top panels of Fig.~\ref{IPR_CT}. This transition is tightly related to the ones recently discussed in~\cite{mirlin_cayley,ioffe2,ioffe3} and will be 
analyzed in full details in a forthcoming paper~\cite{DPRM_CT}.)  

In conclusion, the analysis of the convergence of the LDoS indicate that the Anderson model on the RRG is fully ergodic in the 
whole delocalized phase, ergodicity being eventually restored on a finite energy scale $\eta_c(W)$ (resp., a finite system size $N_c(W)$) 
which becomes exponentially small (resp., exponentially large) as $W_L$ is approached, while the Anderson model on the loop-less
Cayley tree displays a genuine multifractal (non-ergodic) behavior in the whole delocalized phase, as
already discovered in~\cite{garel,mirlin_cayley}.
However, the non-ergodic crossover region observed on the RRG is highly non-trivial:
The apparent multifractal behavior observed on the RRG for $N < N_c (W)$ seems to be controlled by the the multifractal behavior found at the root 
of the Cayley, giving rise to non-trivial desorder-dependent fractal exponents.

\subsection{The level compressibility} \label{sec:chi}

In order to obtain more information on the level and eigenfunctions' statistics of the Anderson model on the RRG and on
the Cayley tree, and to clarify the differences between the two types of 
lattices,
in the remaining part of this section we study two specific observables related to the statistics of energy levels and wave-functions' coefficients, 
which can be easily expressed in terms of the elements of the resolvent matrix, and computed within the BP approach.

Here we start by focusing on the level compressibility, $\chi_N(E)$~\cite{metha} for the number of energy
levels inside the interval $[-E/2,E/2]$, which, as explained below, is a suitable probe to distinguish
between ergodic, localized, and multifractal states~\cite{metz,Alts_chi,chalker_chi,Bogo,mirlin_rev}.
To this aim, we first introduce the number of energy levels inside an energy interval of width $E$ (and centered around zero):
\[
	{\cal N}_N (E) = \int_{-E/2}^{E/2} \sum_{m=1}^N \delta(E^\prime - E_m) \, \textrm{d} E^\prime \, ,
\]
where $E_m$ are the eigenvalues of the Hamiltonian.
The level compressibility is defined as the ratio between the variance of ${\cal N}_N (E)$, characterizing the fluctuations of energy level
within $[-E/2,E/2]$, and its average~\cite{metha}:
\[
	\chi_N(E) = \frac{\overline{({\cal N}_N (E))^2} - \overline{{\cal N}_N (E)}^2}{\overline{{\cal N}_N (E)}} \, ,
\]
where $\overline{\cdots}$ denotes the average over the disorder.

Let us focus on the behavior of $\chi_N(E)$ when the energy interval is measured in units of the mean level spacings: $E = s \delta$.
In the standard ergodic metallic phase, described by the Wigner-Dyson statistics, energy levels strongly repel each other, and the
variance scales as $\overline{({\cal N}_N (E))^2} - \overline{{\cal N}_N (E)}^2 \propto \ln \overline{{\cal N}_N (E)}$~\cite{metha}.
Hence the level compressibility vanishes as $\chi_N(E) \propto \ln N/N$ for large $N$.
Conversely, in the localized phase energy levels are thrown as random points on a line and are described by a Poisson distribution.
Hence $\overline{({\cal N}_N (E))^2} - \overline{{\cal N}_N (E)}^2 = \overline{{\cal N}_N (E)}$ and $\chi_N(E) \to 1$ for $N \to \infty$.
Finally, for non-ergodic multifractal states the variance of the number of energy levels inside an interval should scale linearly with the 
average~\cite{Alts_chi,chalker_chi,Bogo,mirlin_rev}, and $\chi_N(E)$ is expected to converge to a (system-dependent) constant between $0$ and $1$ in the 
large $N$ limit (at least in simplest scenarios).

The level compressibility in the Anderson model on the RRG has been recently studied in the thermodynamic limit in~\cite{metz}.
However, in this case the limit $N \to \infty$ is taken from the start, while the $s \to 0^+$ and $\eta \to 0^+$ 
limits are taken {\it after} 
the thermodynamic limit. As already explained above, this strategy does not allow to detect the existence of the putative
delocalized non-extended states. One should instead study the behavior of $\chi_N (E)$ at {\it finite} $N$, letting  $s$ 
scale as $N^{\sigma}$, with $\sigma \le 0 \le 1$, thereby enabling to scan the statistics of energy levels on all scales, from that
of the mean level spacing ($\sigma = 0$) up to energies of order one ($\sigma = 1$).
This can be easily achieved in the framework of the BP approximation, 
since ${\cal N}_N (E)$ can be expressed in a simple way in terms of the Green's functions defined on the nodes and 
on the edges of the lattice.
The calculation on the RRG, which is carried out in full details in App.~\ref{app:NE}, yield:
\begin{equation} \label{eq:NE1}
\begin{aligned}
{\cal N}_N (E) =& \! \lim_{\eta \to 0^+} \! \bigg \{ \frac{1-k}{2 \pi} \sum_{i=1}^N \big[ \Psi_i ( z_+ ) - \Psi_i (z_- ) \big]\\
	&\qquad + \frac{1}{2 \pi} \sum_{i=1}^N \sum_{j \in \partial i} \big[ \psi_{i \to j} ( z_+ ) - \psi_{i \to j} ( z_- ) \big] \bigg \} \, ,
\end{aligned}
\end{equation}
where $z_{\pm} = \pm E/2 + i \eta$, and the angles $\Psi_{i \to j} (z)$ and $\psi_{i \to j} (z)$ are defined as the phases of 
${\cal G}_i (z)$ and $G_{i \to j} (z)$ respectively, ${\cal G}_i (z) = | {\cal G}_i (z) | e^{i \Psi_i (z)}$, and $G_{i \to j} (z) = | G_{i \to j} (z)| e^{i \psi_{i \to j} (z)}$
(we have chosen here to put the branch-cut in the complex plane along the negative real axis).
A very similar expression can be obtained for the Cayley tree, Eq.~(\ref{eq:NECT}).
In fact, while in the latter case Eq.~(\ref{eq:NECT}) is an exact formula for ${\cal N}_N (E)$, 
one should keep in mind that due to the presence of loops Eq.~(\ref{eq:NE1}) only provides an approximate expression for 
the number of energy levels on RRGs of finite size (which is expected to become asymptotically exact in the $N \to \infty$ limit). 

\begin{figure}
 \includegraphics[angle=0,width=0.48\textwidth]{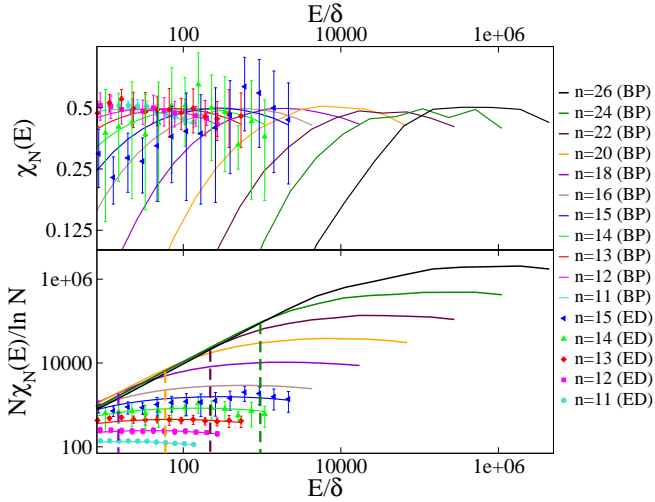}
 \caption{\label{chi-11}
	Top panel: Level compressibility, $\chi_N$, (averaged over many independendent realizations of the on-site disorder
	and of the RRG) 
	plotted as a function of $E/\delta$
	for the Anderson model on the RRG at $W=11$ and for several system sizes $N=2^n$ with $n=11, \ldots, 26$ (and for $c=8$).
	Continuous lines show the results found with the BP approach while full symbols represent the data obtained from
	BP (averaged over $2^{19-n}$ samples only).
	Bottom panel: Same data as above with a rescaling of the $y$-axis as $N \chi_N / \ln N$.
	The vertical dashed lines spot the values of the energy at which the curves corresponding to different sizes deviate 
	from the scaling function.}
 \end{figure}

In order to analyze the scaling properties of the level compressibility $\chi_N(E)$  
we need then to compute the average of ${\cal N}_N (E)$ and its fluctuations over many independent random instances of large but finite 
size, using Eqs.~(\ref{eq:recursion}), (\ref{eq:recursion_final}), and (\ref{eq:NE1}), and investigate their asymptotic behavior in the limit of large $N$.
Hence, three simultaneous limits are involved: $N \to \infty$, $\eta = c \delta \to 0^+$ (with $c=8$ as above), and $E = s \delta \to 0^+$, 
where $\delta = 1/(N \rho (W))$ is the mean level spacings around the middle of the band.
(Note that it does not make much sense to take $s$ smaller than $c$, since the broadening of 
the $\delta$-peaks of the DoS smoothens-out the information on individual 
levels on energy intervals smaller than $\eta$.)

\begin{figure}
 \includegraphics[angle=0,width=0.48\textwidth]{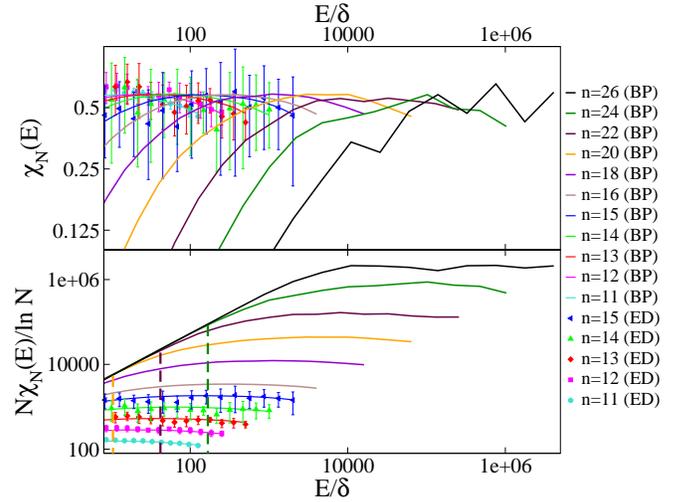}
 \caption{\label{chi-12}
        Top panel: Level compressibility, $\chi_N$, (averaged over many independendent realizations of the on-site disorder and
	of the RRG)
        plotted as a function of $E/\delta$
	for the Anderson model on the RRG at $W=12$ and for several system sizes $N=2^n$ with $n=11, \ldots, 26$ (and for $c=8$).
        Continuous lines show the results found with the BP approach while full symbols represent the data obtained from
        BP (averaged over $2^{19-n}$ samples only).
        Bottom panel: Same data as above with a rescaling of the $y$-axis as $N \chi_N / \ln N$.
        The vertical dashed lines spot the values of the energy at which the curves corresponding to different sizes depart
        from the scaling function.}
 \end{figure}

As far as the existence of the putative non-delocalized phase is concerned, the scaling behavior of the level compressibility
on the scale of the mean level spacing only [i.e., for $s$ of $O(1)$] might be uninformative:
Consider, for instance, the model of Ref.~\cite{kravtsov} of the Rosenzweig-Porter type, where an intermediate mixed phase can
be explicitely realized in some region of the parameter space. It can be shown that in such phase the level statistics on the scale of the 
mean level spacing is still described by the GOE ensemble, whereas a crossover to Poisson statistics takes place on a scale $N^{D_2-1}$ which 
goes to zero with $N$ but stays much larger than $\delta$.
In order to be able to describe this situation, we let $s$ be equal to $s = c N^\sigma$,
and consider seveal values of $\sigma \in [0,1)$. This allows to probe the statistics of energy levels at all scales $E \propto N^{\sigma-1}$
spanning the whole energy range from the scale of the mean level spacing ($\sigma = 0$) up to energies of $O(1)$ ($\sigma \to 1$).

The results for $\chi_N (E)$ for the Anderson model on the RRG are plotted in Figs.~\ref{chi-11} and \ref{chi-12} for $W=11$ and
$W=12$ respectively, as a function of the energy measured in units of the mean level spacing $\delta$, for several system sizes, $N=2^n$, 
with $n=11, \ldots , 26$.
The level compressibility has been averaged over many independent realizations of the on-site disorder and
of the RRG.
From the top panels we notice that at large enough energy (and/or small enough $N$), 
$\chi_N (E)$ seems to approach a constant value between zero and one ($\chi \approx 0.49$ for $W=11$ and 
$\chi \approx 0.57$ for $W=12$), which is a typical signature of non-ergodic multifractal states. However, 
when the energy is decreased below a certain value, $\chi_N(E)$ departs from the plateau value and decreases to zero.
The energy at which $\chi_N(E)$ reaches the plateau grows proportionally to $N$ as the system size is increased.
Hence, if the system size is too small (i.e., $N < N_c(W)$) one is not able to observe the departure from the plateau
and the system behaves as if it was in a genuine non-ergodic phase, with a well defined value of $\chi \in (0,1)$.
We also show the data obtained from ED (filled symbols) up to the largest available system size, $N=2^{15}$, 
(averaged however over much fewer samples). They are in reasonably good agreement within the numerical accuracy
with the BP results.

In the bottom panel we plot the rescaled level compressibility, $N \chi_N (E) / \ln N$, which should 
collapse onto a $N$-independent scaling function in the limit of large sizes if the Wigner-Dyson statistics is recovered. 
This is precisely what we observe in the bottom panels, which exhibit a nice collapse for small enough energies and large enough sizes.
The values of the energy at which the curves corresponding to different $N$ deviate from the scaling function are spotted
as vertical dashed line, and are found to scale proportionally to $N$ for large enough sizes.
This behavior indicates that, provided that $N$ is sufficiently large, ergodicity and GOE statistics are eventually recovered 
in the delocalized phase of the Anderson model on the RRG on an energy scale which remains finite (and which vanished exponentially at $W_L$).

\begin{figure}
 \includegraphics[angle=0,width=0.48\textwidth]{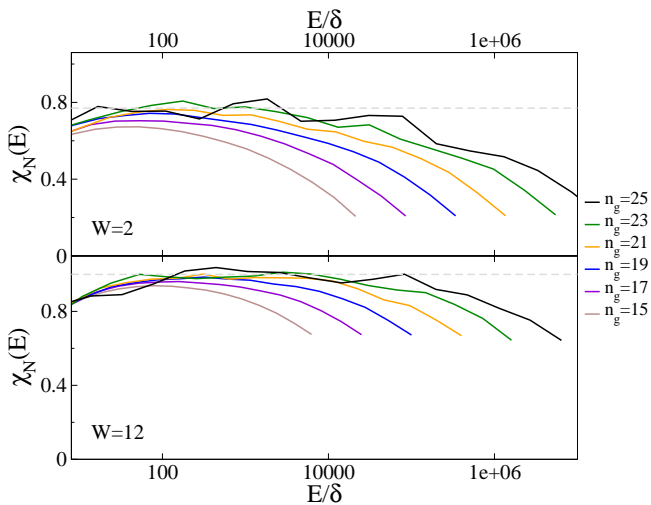}
 \caption{\label{chi-CT}
        Level compressibility, $\chi_N$, (averaged over many independendent realizations of the on-site disorder)
        plotted as a function of $E/\delta$
	for the Anderson model on Cayley trees of $n_g$ generations (with $n_g = 15, \ldots, 25$) at $W=2$ (top panel) and $W=12$ (bottom panel)
	for $c=8$. The horizontal gray dashed lines show the approximate plateau value of $\chi$,
        associated to sub-Poissonian statistics.}
 \end{figure}

The situation is drastically different on the Cayley tree, as shown in Fig.~\ref{chi-CT}.
We indeed observe that, when the number of generations $n_g$ of the tree is increased, the level compressibility approaches 
asymptotically a function which is roughly constant 
and which stretches to larger and larger values of the energy as the
system size is increased. This is a clear signature of multifractal non-ergodic states characterized by sub-Poissonian 
statistics on all energy scales~\cite{Alts_chi,Bogo,mirlin_rev}. 
The plateau value of $\chi$ (green dashed lines)
increases as $W$ is increased, and is already large at small disorder (e.g., $\chi \approx 0.77$ for 
$W=2$), and is very close to unity
at moderate disorder strength ($\chi \approx 1$ at $W=12$), 
compatible with the localization of wave-functions close to the boundary of the tree.

\subsection{The overlap correlation function} \label{sec:K2}

Another very useful probe of the statistics of the eigenfunctions which allows to distinguish between ergodic, localized, and multifractal
states is provided by the overlap correlation function between eigenstates at different energy levels~\cite{kravtsov,krav_K2,chalker_K2,thouless,mirlin_rev}, 
defined as:
\begin{equation} \label{eq:corr}
	K_2 (E) = \frac{ N \sum\limits_{i} \sum\limits_{m,m^\prime} | \braket{i}{m} |^2 | \braket{i}{m^\prime} |^2 \delta 
	\big[ E - (E_{m} - E_{m^\prime}) \big]}
	{\sum\limits_{m,m^\prime} \delta \big[ E - (E_{m} - E_{m^\prime}) \big]} \, ,
\end{equation}
where $\braket{i}{m}$ is the amplitude of the eigenvector $\ket{m}$ on site $i$.

For eigenfunctions of GOE matrices $K_2 (E) = 1$ identically, independently on $E$ on the entire spectral band-width.
In the standard (ergodic) metallic phase 
$K_2(E)$ has a plateau at small energies, $K_2(E) \simeq q_2$ for $E < E_{Th}$, followed by
a fast-decay which is described by a power-law, $K_2(E) \sim E^{-\gamma}$, with a system-dependent exponent~\cite{chalker_K2}.
The height of the plateau is larger than one, which implies an enhancement of correlations compared to the
case of independently fluctuating Gaussian wave-functions.
The Thouless energy, $E_{Th}$, which separates the 
plateau from the power-law decay stays finite in the thermodynamic limit and extends to larger energies as one goes deeply
into the metallic phase, and corresponds to the energy range over which GOE-like correlations establish~\cite{thouless}.

The behavior of the overlap correlation function for multifractal wave-functions is instead drastically different, 
as shown in~\cite{kravtsov}: The plateau is present only in a narrow energy interval $E < E_{Th} \sim \delta N^{D_2}$
which shrinks to zero in the thermodynamic limit as $N^{D_2-1}$, while its height grows $N^{1-D_2}$.
This can be interpreted recalling that multifractal wave-functions typically occupy a fraction $N^D$ of the total sites, 
which implies the existence of an energy scale, $E_{Th}$, which decreases with $N$ but stays much larger than the
mean level spacing, beyond which eigenfunctions poorly overlap with each other and the statistics is no longer GOE.

\begin{figure}
 \includegraphics[angle=0,width=0.48\textwidth]{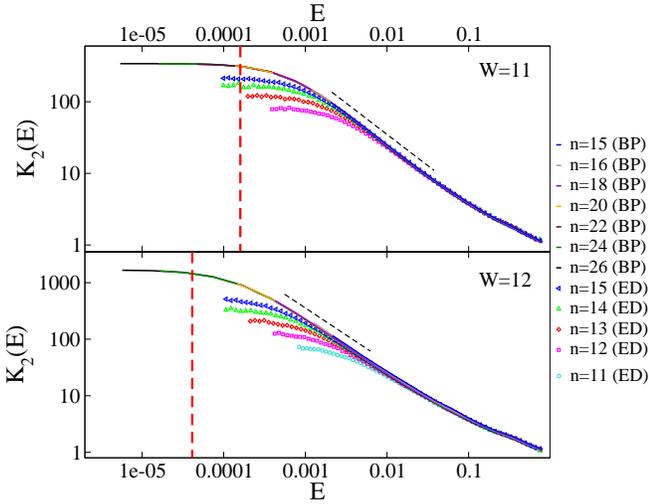}
 \caption{\label{K2RRG}
	Overlap correlation function $K_2(E)$ (averaged over many independendent realizations of the Hamiltonian)
        plotted as a function of $E$
        for the Anderson model on RRGs of $2^N$ sites (with $n = 11, \ldots, 26$) at $W=11$ (top panel) and $W=12$ (bottom panel)
	for $c=8$.
	Continuous curves show the results obtained within the BP approximation and symbols correspond to the data
	obtained using ED up to $N=2^{15}$. The vertical red dashed lines spot the position of the Thouless energy.
	The dashed black lines represent
	the power-law decay from the plateau with exponents $\gamma \approx 1$ independently on the disorder~\cite{ioffe3}. }
 \end{figure}

For any given random instance of the Hamiltonian, 
the overlap correlation function~(\ref{eq:corr}) can be easily expressed in terms of the Green's functions
computed at energies $\pm E/2$ as:
\[
	K_2 (E) = \lim_{\eta \to 0^+} \frac{N \sum_i {\rm Im}{\cal G}_i (-E/2) \, {\rm Im}{\cal G}_i (E/2)}
	{\sum_i {\rm Im}{\cal G}_i (-E/2) \sum_i {\rm Im}{\cal G}_i (E/2)} \, .
\]
In order to determine the scaling properties of the overlap correlation function, 
we have computed the average of $K_2(E)$ over many independent realizations of the disorder for the
Anderson model on the RRG and on the Cayley tree, using the expression above where the Green's functions
are evaluated at the fixed point solution of the BP equations, and for energy differences varying from the scale of the mean level 
spacing up to energy differences of $O(1)$.

The results for the RRG are plotted in Fig.~\ref{K2RRG} for $W=11$ and $W=12$, showing that the $N$-dependence of 
$K_2(E)$ saturates for large enough $N$ and that the curves converge to a $N$-independent limiting function
characterized by a plateau at small energy followed by a fast decrease [$K_2(E) \sim (E_{Th}/E)^\gamma$] 
at large energy corresponding to
the onset of level repulsion (with $\gamma \approx 1$ independently of $W$~\cite{ioffe3}). 
The crossover from the plateau to the power-law decay takes place on the
energy scale $E_{Th}$ (vertical red dashed lines), which stays finite in the thermodynamic limit and 
represents the width of the energy band within which GOE-like correlations are established~\cite{thouless}.
This behavior is very similar to the one found in the metallic phase of the $3d$ Anderson model close to the critical
point. In particular, the fact that the plateau survives in the $N \to \infty$ limit and extends to larger energies as one goes deeply into the conducting phase 
is a clear signature of ergodic states~\cite{krav_K2,chalker_K2}.
However, the fact that its value is much larger than one is
an apparent manifestation of the enhancement of correlations and of the fact that wave-functions show significant
deviations from uncorrelated Gaussian random variables.
We again observe an excellent agreement between the results obtained using the BP approximation and EDs
(note, however, that the BP approximation does not allow to access energies smaller than the broadening of
the energy levels, $c \eta$, for the reasons explained above). Nevertheless, at $W=11$ and $W=12$, deep into the 
non-ergodic-like crossover regime, the
largest system sizes via ED are too small to allow to observe the convergence of $K_2(E)$.

The Thouless energy is found to be proportional to the energy scale $\eta_c(W)$ below which the probability distribution of the local DoS converges
to a stable non-singular distribution (see Fig.~\ref{fit_length}), and thus vanishes exponentially at $W_L$.
Moreover, $E_{Th}$ turns out to coincide (within our numerical accuracy) with the energy scale below which the Wigned-Dyson 
asymptotic scaling of the level compressibility is recovered (vertical dashed lines of Figs.~\ref{chi-11} and \ref{chi-12}), 
indicating that the energy band within which the statistics of energy levels is described by the Wigner-Dyson
statistics coincides with
the one over which wave-functions correlations are GOE-like and $K_2(E)$ has a plateau.

\begin{figure}
 \includegraphics[angle=0,width=0.48\textwidth]{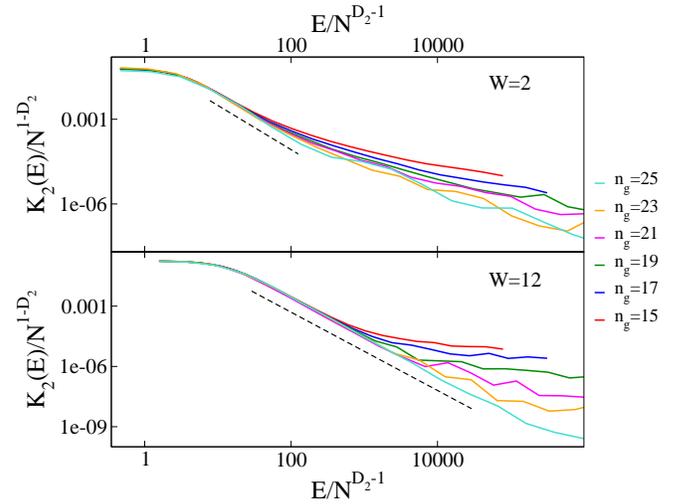}
 \caption{\label{K2CAY}
        Overlap correlation function $K_2(E)$ (averaged over many independent realizations)
        plotted as a function of $E$ for the Anderson model on Cayley trees of $n_g$ generations (with $n_g = 15, \ldots, 25$) 
	at $W=2$ (top panel) and $W=12$ (bottom panel).
	$K_2(E)/N^{1-D_2}$ for different $N$ collapse (for $N$ large enough) onto the same curve as a function of $E/E_{Th} \propto E/N^{D_2-1}$ (with $D_2 \approx 0.05$ for $W=2$ and $D_2 \approx 0.002$ for $W=12$, see Fig.~\ref{D2multi}). 
	The dashed black lines represent
	the power-law decay from the plateau with exponents $\gamma \approx 1.35$ for $W=2$ and $\gamma \approx 1.95$ 
	for $W=12$.}
 \end{figure}

The situation on the Cayley tree is completely different. In this case, as shown in Fig.~\ref{K2CAY}, $K_2(E)$ presents all the distinctive features typically
observed for multifractal states: 
the Thouless energy decreases with the system size as $\delta N^{D_2} \propto N^{D_2-1}$ whereas the height of the plateau grows as
$N^{1-D_2}$. The curves of $K_2(E)/N^{1-D_2}$ for different $N$ collapse (for large enough $N$ and small enough energies) onto the same curve once the energies
are rescaled by $E_{Th}$. 
In fact, as discussed above, the value of $D_2$ is actually very close to zero at moderate disorder strength
($D_2 \approx 0.002$ at $W=12$) 
and is already very small at weak disorder ($D_2 \approx 0.04$ for $W=2$).
Note that the power-law decay from the plateau, $K_2 (E) \sim (E_{Th}/E)^\gamma$, observed on the Cayley tree
is quite 
different with respect to the RRG: We find that the exponent $\gamma$ is
greater than one and slowly increases with $W$ ($\gamma \approx 1.35$ at $W=2$ and $\gamma \approx 1.95$ at $W=12$). 
Interestingly, in the region where the fractal exponents $D_1$ and $D_2$ are close to zero ($W \gtrsim 10$,
see Figs.~\ref{Dmulti} and~\ref{D2multi}) the value of the exponent $\gamma$  
is very close to $\gamma \approx 2$, which is the same found in the whole delocalized non-ergodic phase
of the random matrix model of the Rosenzweig-Porter type of Ref.~\cite{kravtsov,facoetti}.

\section{Recap of the main results, conclusions, and perspectives} \label{sec:conclusions} 

In this paper we have studied the Anderson model 
on two different kinds of Bethe lattices, the RRG and the loop-less Cayley tree,
focusing in particular on the ergodic properties of the delocalized phase on these
two lattices. Our analysis is based on a novel approach which consists in solving the 
iteration relations for the Green's functions directly on random instances of large but finite
sizes. 
We start this section by  giving below a sketchy summary of the main results.

\vspace{6pt}

\noindent {\bf 1) Exact diagonalization on the RRG: Characteristic crossover scale.}\\
In Sec.~\ref{sec:numerics} we have presented 
an accurate numerical analysis of several observables and probes associated to
level and eigenfunction statistics that display 
different universal behaviors in the ergodic and non-ergodic regimes
(such as the ratio of adjacent gaps, the overlap between eigenvectors corresponding to 
subsequent eigenvalues, the IPR, the wave-functions' support set, and their spectrum of
fractal dimensions). We performed
EDs on the delocalized side of the Anderson transition
on RRGs of size $N$ from $2^6$ to $2^{15}$.
Our results clearly show the existence of a characteristic system size governing
finite size effects, $N_c (W)$, as already observed in~\cite{mirlin,levy,ioffe1}, 
which diverges much faster than a power-law approaching the localization transition
(as predicted by the supersymmetric analysis~\cite{fyod})
and is already very large far from it.
The most important observation is that the behavior of {\it all} the considered observables,
both those associated to the statistics of energy levels on the scale of the mean level 
spacings, and those related to the statistics of wave-functions,
is governed by the correlation volume $N_c (W)$ (see Fig.~\ref{nc}), suggesting that the crossover from
Poisson statistics and multifractal wave-functions to GOE statistics and ergodic wave-functions
occurs concomitantly.

\vspace{6pt}

\noindent {\bf 2) BP solution: Convergence of the local density of states and fractal exponents.}\\
In Sec.~\ref{sec:BP} we discussed the results found computing the BP solution of the self-consistent iteration equations for the Green's functions of the
Anderson model on the RRG and on the Cayley tree on very large but finite instances of size $N$ from $2^{11}$ 
to $2^{29}$ sites.
In Sec.~\ref{sec:BPtest} we have shown that the results obtained using the BP approximation on the RRG are in excellent agreement with
the exact solution obtained from ED (up to the largest system sizes accessible via ED, $N=2^{15}$),
provided that the imaginary regulator $\eta$ is of the order of the mean level spacing, i.e., $\eta = c \delta$, 
with $\delta = 1/ ( N \rho(W))$ [where $\rho(W)$ is the average DoS at the center of the band].
We show in particular that the BP solution provides a tight and controlled approximation not only for average and/or
global quantities, but also for local observables, and accounts accurately for sample to sample and spatial fluctuations.
(Note that  BP is exact on the Cayley tree due to the absence of loops.)
    
In Sec.~\ref{sec:BPLDoS} we focused on the probability distribution of the LDoS obtained within the BP approach on the RRG, and showed that
the dependence on the system size of $P({\rm Im} {\cal G})$ saturates for large enough sizes (i.e., $N \gg N_c (W)$ or, equivalently, for $\eta$ smaller
than a disorder-dependent energy scale $\eta_c(W)$ which stays finite in the delocalized phase and vanishes exponentially at $W_L$), and 
convergence to a stationary, size independent, stable, non-singular, probability distribution is observed (at least up to the largest
accessible disorder strength $W \lesssim 13.5$).
Interestingly, the crossover scale $N_c(W)$ obtained from the convergence of
the LDoS within the BP approach, Eq.~(\ref{eq:Nc}), accounts very well for the scale above which ergodic behavior emerges (see Fig.~\ref{nc}).

Conversely, we observed that the Anderson model on the Cayley tree displays a genuine multifractal, non-ergodic behavior at
all scales in the whole delocalized phase, in agreement with~\cite{garel,mirlin_cayley}.
We computed the fractal exponents $D_1$ and $D_2$  associated to the spectral statistics, which exhibit a non-trivial dependence on the position inside the 
tree~\cite{mirlin_cayley,DPRM_CT}, and we showed that the apparent non-ergodic features observed on the RRG for $N<N_c$ seems to be controlled by the multifractal properties of the region close to the root of the Cayley tree at the same disorder strength. 

\vspace{6pt}

\noindent {\bf 3) Level compressibility and overlap correlation function.}\\
In Secs.~\ref{sec:chi} and \ref{sec:K2} 
we focused on two spectral probes, such as the level compressibility $\chi_N (E)$~\cite{metha} and the overlap correlation function $K_2(E)$~\cite{krav_K2},  
associated respectively with the statistics of level spacings and eigenfunctions 
that display very different scaling behavior in the delocalized, localized and intermediate mixed 
phase~\cite{kravtsov,metz,mirlin_rev,krav_K2,Alts_chi,chalker_chi,Bogo,chalker_K2,thouless}.
These observables can be easily expressed in terms of the Greens' functions obtained from the BP solution of the Anderson model 
on the RRG and on the Cayley tree. 
Their analysis on the RRG reveal the existence of an energy scale, $E_{Th} (W)$, which remains finite in the whole delocalized phase, corresponding to the window in energy within which the Wigner-Dyson level statistics is recovered
and eigenfunctions exhibit GOE-like correlations, corresponding to a size-independent plateau of $K_2 (E)$ at small energy separation~\cite{thouless}. 
Such energy scale vanishes exponentially fast approaching $W_L$ and is in fact proportional to $N_c^{-1}$.
Hence, for $N<N_c(W)$ the mean level spacing is larger than $E_{Th} (W)$ and the system
looks like as if it were in an intermediate non-ergodic delocalized phase.

Conversely, on the Cayley tree the behavior of $\chi_N (E)$ and $K_2(E)$ is fully consistent with the existence of genuinely multifractal states in the whole delocalized phase
(with localization of the wave functions close to the boundary of the tree). 
In particular, energy levels on the Cayley tree exhibit a sub-Poissonian statistics (in fact, very close to Poissonian already very far from $W_L$), while the analysis of
eigenfunctions' correlations show the existence of an energy scale which decreases with $N$ (as $N^{D_2-1}$) but stays larger than the mean level spacing,
which is the hallmark of non-ergodic extended states.

\vspace{6pt}

All in all, the results presented in this paper support in a coherent way the idea that the Anderson model on the RRG becomes fully ergodic in the whole delocalized phase: 
ergodicity and GOE statistics are 
eventually recovered in the thermodynamic limit in the whole extended phase, implying that 
the GOE-ergodic/Poisson-non-ergodic transition of the energy levels and eigenvectors
is concomitant with Anderson localization, in agreement with the recent results of~\cite{mirlin,levy,lemarie} and with the predictions of~\cite{SUSY,fyod} based
on supersymmetric field theory.
Nonetheless, ergodicity establishes on a system size (resp., energy scale) which becomes exponentially large (resp., small) 
as the localization transition is approached,
and exceeds the system sizes accessible via ED well
before the localization transition,
resulting in a very wide crossover region in which the system looks as if it were in
a mixed (delocalized but non-ergodic) phase for all practical purposes, i.e. on finite but large 
length and time scales (volumes smaller than $N_c(W)$ and times smaller than $\hbar/E_{Th} (W)$).

Furthermore, the apparent non-ergodic-like crossover region observed on the RRG for $N<N_c$ has highly non-trivial properties,
and is characterized by a set of effective disorder-dependent fractal exponents which are independent on $N$ in a broad range of
system sizes. Such apparent multifractal behavior seems to be controlled 
by the one of the root of the Cayley tree at the same disorder strength.
Indeed, a genuine non-ergodic extended phase is found in the Anderson model on the loop-less Cayley tree in the whole delocalized side, as predicted in~\cite{garel, mirlin_cayley}. The properties of such phase will be discussed in more details in a forthcoming paper~\cite{DPRM_CT}

On the basis of the analogy between Anderson localization on Bethe lattices and 
Many-Body Localization~\cite{A97,BAA,jacquod,wolynes,scardicchioMB}, 
these phenomena might play a very important role and lead to highly non-trivial behaviors 
in the delocalized phase of 
many-body interacting disordered systems exhibiting MBL~\cite{dinamica,DPRM_CT}.




\vspace{6pt}

Given the difficulty of the questions we are addressing, it is natural to dwell about possible limitations
of our analysis. For instance, there is the possibility that for some reason the BP approach
starts to fail in some region of the parameters space, and in particular within the putative delocalized non-ergodic
phase, $W \ge W_E$ and $N$ very large.
However, besides the fact that an excellent agreement between the BP approximation and ED results is found for all observables and probes considered and that
BP passed successfully all the numerical tests of Sec.~\ref{sec:BPtest},
there are no exemples in the literature of other models where something similar might happen.
On the contrary, the BP approximation is expected on general grounds to improve as $N$ is increased~\cite{mezard}.
Yet, although there is a rigorous proof of the convergence of the BP solution for the Anderson model on the RRG in the large $N$ limit~\cite{bored}, there is no rigorous estimate of the error at large but finite $N$.
It would be very interesting in this respect to characterize in a quantitative way the convergence of both local and average observables obtained from the BP approximation. In standard statistical mechanics models one generally finds that the finite-size corrections of BP for global quantities, such as, e.g., the free-energy, are of order $O(1/N)$ (in the replica-symmetric phase)~\cite{BPconvergence}. Here instead our numerical results
suggest that, up to the moderately large size accessible via ED, global observables approach their exact values as $1/\sqrt{N}$. Further work is necessary to obtain more
definite conclusions.

Another point worth mentioning is that all results discussed above are valid for $\eta > \delta$, where the simultaneous limits $N \to \infty$
and $\eta \propto 1/N \to 0^+$ are taken.
Recent studies of the LDoS on the delocalized side of the Anderson model on the RRG seem to suggest
that its statistical properties might be unusual in the regime $\eta \ll \delta$~\cite{ioffe_private}.
As discussed above, the BP approach is not applicable to this situation and here we only focused on the more
standard case $\eta > \delta$.

Another related interesting perspective would be to banchmark the BP framework onto the random
matrix models of the Rosenzweig-Porter type of Ref.~\cite{kravtsov,facoetti}, which is characterized by a
whole region of the parameter space where wave-functions are delocalized but truly multifractal.
Preliminary results (which will be discussed in a forthcoming work~\cite{future}) 
indicate that in this case BP is able to detect correctly the 
presence of the delocalized non-ergodic states.

\vspace{6pt}

\appendix

\section{Multifractality} \label{app:multi}
In this appendix we give more information and details on the computation of the spectrum of fractal dimensions of wave-functions coefficients.
In order to obtain $f_N (\alpha)$,
we have computed the average of the moments $\langle \Upsilon_q (n) \rangle$, for different system sizes
$N = 2^n$, with $n$ from $6$ to $15$,
and for $400$ different values of $q$ in the interval $(-3,5)$.
Data are averaged over (at least) $2^{22-n}$ samples, and over $1/8$ of the eigenstates around the middle of the band.
For each value of the disorder strength $W$, $\tau_N (q)$ is obtained as (minus) the derivative
of the logarithm of the moments with respect to the logarithm of the system size, 
which can be approximately evaluated as:\footnote{Note that we have performed an {\it annealed} 
computation (logarithm of the average) instead of the {\it quenched} one (average of the logarithm).
One can show that the spectrum of fractal dimensions obtained using the two definitions coincide as far as $f(\alpha)>0$,
{\it i.e.}, in the whole support $\alpha \in (\alpha_-, \alpha_+)$.}
\begin{equation} \label{eq:MFtau}
\tau_N (q) = - \, \frac{ \ln \langle \Upsilon_q (n) \rangle - \ln \langle \Upsilon_q (n - \delta n) \rangle}
{\delta n \ln 2} \, .
\end{equation}
We then have computed $\alpha_N (q)$ as the derivatives of
$\tau_N (q)$ with respect to $q$:
\begin{equation} \label{eq:MFalpha}
\alpha_N(q) = \frac{\tau_N (q + \delta q) - \tau_N (q)}{\delta q} \, .
\end{equation}
For simplicity, in most of the cases we have chosen 
$\delta n = 1$,\footnote{except for $W=5$, $W=10$, and $W=13$ where we have considered smaller values
of $\delta n$ in order to obtain more precise results.}
and we have used $\delta q = 5 \cdot 10^{-5}$.
Finally, we evaluate numerically the Legendre transform as
$f_N(\alpha_N) = q \alpha_N (q) - \tau_N (q)$, where $\tau_N (q)$ and $\alpha_N$ are given by Eqs.~(\ref{eq:MFtau})
and (\ref{eq:MFalpha}).
 
 \begin{figure}
 \includegraphics[angle=0,width=0.44\textwidth]{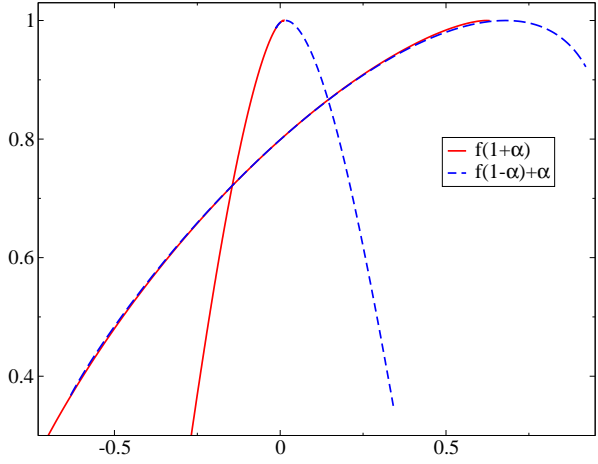}
 \caption{\label{Multi-Symm} 
Verification of the symmetry relation of Eq.~(\ref{eq:MFsymm}), for $W=5$ and $W=13$ and for $n=10$ (coinciding red continuous curves and 
blue dashed curves).
Similar plots are found for other values of $W$ in the delocalized phase and for other system sizes.}
 \end{figure}

As demonstrated in~\cite{scardicchio1,scardicchio2}, in the region of extended states the spectrum of fractal dimensions
should obey the following symmetry relation:
\begin{equation} \label{eq:MFsymm}
f(1+\alpha) = f(1 - \alpha) + \alpha \, .
\end{equation}
In order to check the accuracy of our numerical procedure, 
in Fig.~\ref{Multi-Symm} we verify that the non-trivial symmetry~(\ref{eq:MFsymm}) is indeed nicely 
fulfilled for $f_N (\alpha)$
for two values of the disorder in the delocalized phase ($W=5$ and $W=13$) and for $N=2^{10}$. Similar outcomes
are found for different values of $W$ in the extended regime and for other values of $N$.
 
In the following we will focus in particular on the $N$-dependence of four specific points of the singularity spectrum: 
the most probable value $\alpha_m$ where $f_N(\alpha)$ reaches its maximum, $f_N (\alpha_m) = 1$;
the point $\alpha_1$ (associated to $q=1$) where $f_N (\alpha_1) = \alpha_1$, and $f_N^\prime(\alpha_1) = 1$;
the lower edge of the support of $f_N(\alpha)$, $\alpha_-$;
the point $\alpha_{\rm cross}$ where the spectra of fractal dimensions for two subsequent system sizes cross.

\begin{figure}
 \includegraphics[angle=0,width=0.46\textwidth]{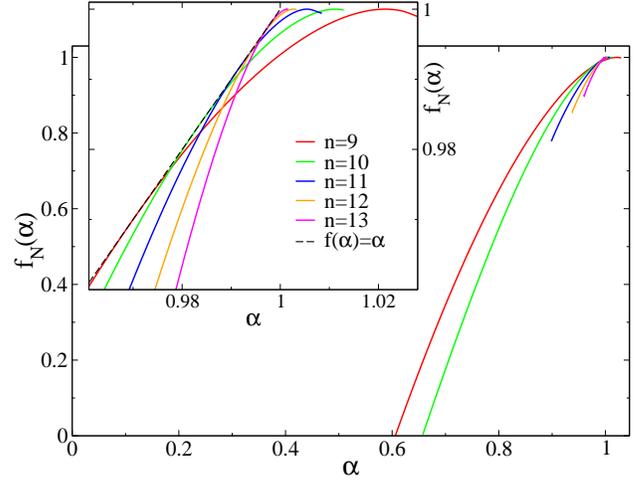}
 \caption{\label{Multi-5} 
Spectrum of fractal dimensions $f_N (\alpha)$ for $W=5$ and for different system sizes $N=2^n$ with $n$ from $9$ to $13$.
The inset shows a zoom of the same curves in the region close to $\alpha = 1$. The straight line $f(\alpha) = \alpha$ (black dashed line) is
tangent to $f_N (\alpha)$ in $\alpha_1$.}
 \end{figure}

In Fig.~\ref{Multi-5} the singularity spectrum is plotted for $W=5$ and for 
several system sizes $N=2^n$ with $n$ from $9$ to $13$ (the inset shows a zoom of the same curves 
in the region close to $\alpha = 1$). 
One clearly observes that the support of $f_N (\alpha)$ shrinks as $N$ is increased.

\begin{figure}
 \includegraphics[angle=0,width=0.46\textwidth]{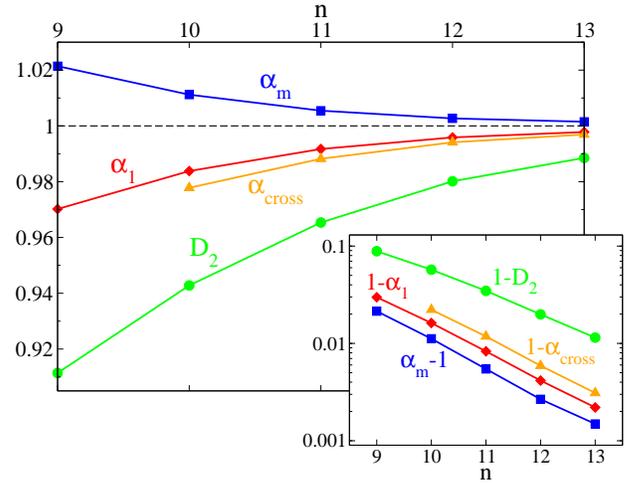}
 \caption{\label{Points-5} 
 Main panel: $D_2$, 
 $\alpha_m$, $\alpha_1$, and $\alpha_{\rm cross}$
  as a function of $n = \log_2 N$.
 Inset: $D_2$, $\alpha_m$, $\alpha_1$, and $\alpha_{\rm cross}$ approach $1$ exponentially in $n$ on the
 same characteristic scale.}
 \end{figure}

From Fig.~\ref{Multi-5} we determine the value of $\alpha_1$ (where $f_N (\alpha)$ is tangent to the straight line $f(\alpha)=\alpha$, as 
shown in the inset), $\alpha_m$, 
and $\alpha_{\rm cross}$ 
as a function of (the logarithm of) $N$.
In Fig.~\ref{Points-5} we show that $\alpha_1$, $\alpha_m$, and $\alpha_{\rm cross}$ all approach $1$ 
as the system size is increased. 
In the same figure we also plot the $n$-dependence of the exponent $D_2$ describing the scaling of the typical value
of the IPR with the system size for $W=5$, introduced in sec.~\ref{sec:IPR-SS}.
As shown in the inset, $\alpha_1$, $\alpha_m$, $\alpha_{\rm cross}$, and $D_2$ all
tend to $1$ exponentially in $n$ on the same characteristic scale.
These results confirm that $f_N (\alpha)$ converges to a $\delta$-function for large $N$, 
$\lim_{N \to \infty} f_N (\alpha) = \delta (\alpha - 1)$,
corresponding to the recovery of full ergodicity.

\begin{figure}
 \includegraphics[angle=0,width=0.46\textwidth]{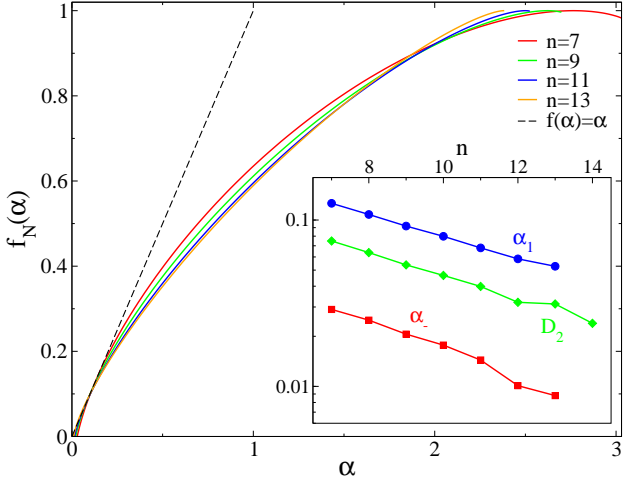}
 \caption{\label{Multi-18} 
 Main panel: Spectrum of fractal dimensions $f_N (\alpha)$ for $W=19$ and for different system sizes 
 $N=2^n$ with $n$ from $7$ to $13$.
 Inset: Behavior of $D_2$, $\alpha_1$, and $\alpha_-$, showing that they all decrease 
 exponentially to zero as a function of $n$ on the same characteristic scale.}
 \end{figure}

Conversely, in the localized regime (see the main panel of Fig.~\ref{Multi-18} for $W=19$), the spectrum of fractal dimensions
gets broader as the system size is increased and
shows a shape which is reminiscent of the triangular form typically observed in the insulating phase.
As a verification, in the inset we focus on the behavior of $\alpha_1$ and $\alpha_-$ as a function of $n$. We also plot the
$n$-dependence of the exponent $D_2$ describing the scaling of the typical value of the IPR with $N$. 
One finds that $\alpha_1$, $\alpha_-$ and $D_2$ all seem to vanish exponentially with
$n$---as expected for localized states---on the same characteristic scale.

\section{Functional iteration relation for the probability distributions of the Green's functions on the Cayley tree} \label{app:cayley}

Due to the presence of the boundary, the sites of the Cayley tree are not translationally invariant even after averaging over the diagonal disorder of the Hamiltonian.
In order to obtain the functional iteration equations for the probability distributions of the Green's functions, one needs then to distinguish their position 
inside the tree, by taking into account their distance from the root.
This can be done by introducing at each generation $\ell$ the probability distributions of two types of cavity Green's functions,
$U_\ell(\overleftarrow{G})$ and $V_\ell(\overrightarrow{G})$ defined, respectively, in absence of the
edge with a site of the previous or the next generation. These functions must satisfy the following functional equations:
\begin{displaymath} 
\begin{split}
	U_\ell (\overleftarrow{G}) &= \! \int \! \textrm{d}p(\epsilon) \prod_{i=1}^k \textrm{d} U_{\ell+1} (\overleftarrow{G}_i) \, 
	\delta \! \left ( \! \overleftarrow{G}^{-1} \! + \epsilon + z + \sum_{i=1}^k \overleftarrow{G}_i \! \right) \, , \\
	V_{\ell^\prime} (\overrightarrow{G}) &= \! \int \! \textrm{d}p(\epsilon) \prod_{i=1}^{k-1} \textrm{d} U_{\ell^\prime + 1} (\overleftarrow{G}_i) \, 
	\textrm{d} V_{\ell^\prime - 1} (\overrightarrow{G}_0) \\
	& \qquad \qquad \qquad \times \delta \! \left ( \! \overrightarrow{G}^{-1} \! + \epsilon + z + \sum_{i=1}^{k-1} \overleftarrow{G}_i 
	+ \overrightarrow{G}_0 \! \right) \, ,
\end{split}
\end{displaymath}
with $\ell = 0, \ldots , n_g$ and $\ell^\prime = 1, \ldots , n_g-1$, with the initial condition at the boundary:
\begin{displaymath} 
\begin{split}
	U_{n_g} (\overleftarrow{G}) &= \! \int \! \textrm{d}p(\epsilon) 
        \delta \! \left ( \! \overleftarrow{G}^{-1} \! + \epsilon + z \! \right) \, ,
\end{split}
\end{displaymath}
and with the prescription that $V_{0} (G) \equiv U_0 (G)$.
from the equations above, one can finally obtain the probability distributions of the Green's functions at any generation of the tree:
\begin{displaymath} 
\begin{split}
	R_{\ell} ({\cal G}) &= \! \int \! \textrm{d}p(\epsilon) \prod_{i=1}^{k} \textrm{d} U_{\ell + 1} (\overleftarrow{G}_i) \, 
        \textrm{d} V_{\ell - 1} (\overrightarrow{G}_0) \\
        & \qquad \qquad \qquad \times \delta \! \left ( \! {\cal G}^{-1} \! + \epsilon + z + \sum_{i=1}^{k} \overleftarrow{G}_i 
        + \overrightarrow{G}_0 \! \right) \, , \\
	R_0 ({\cal G}) &= \! \int \! \textrm{d}p(\epsilon) \prod_{i=1}^{k+1} \textrm{d} U_{1} (\overleftarrow{G}_i) \, 
        \delta \! 
	\left ( \! {\cal G}^{-1} \! + \epsilon + z + \sum_{i=1}^{k+1} \overleftarrow{G}_i \! \right) \, ,
\end{split}
\end{displaymath}
with $\ell = 1, \ldots, n_g$.
Note that deep in the bulk of the tree, in the limit $n_g \to \infty$ at finite $\eta$, the probability distributions becomes 
$\ell$-independent, we recover the
functional equations~(\ref{eq:PGcav}) and (\ref{eq:PG}) found for infinite RRGs.
However, if one consider the simultaneous limits $N \to \infty$ and $\eta \propto 1/N \to 0^+$, the fixed point
of Eqs.~(\ref{eq:PGcav}) and (\ref{eq:PG}) is never reached and is immaterial as far as the spectral statistics is concerned.

\section{Calculation of the number of energy levels ${\cal N}_N (E)$} \label{app:NE}

In this appendix we show how to express the number of energy levels inside the interval $[-E/2,E/2]$, ${\cal N}_N (E)$, in terms of the 
Green's functions 
and the cavity Green's functions 
defined within the BP approach.
In order to do this, one can proceed in two equivalent ways, either 
using the representation of the Heaviside step function $\theta(x)$ (for $x \in {\mathbb R}$) in terms of the discontinuity 
of the complex logarithm along the negative real axis, $\theta (x) = \frac{1}{2 \pi i} \lim_{\eta \to 0^+} [\ln (x + i \eta) - \ln(x - i \eta)]$, 
as done in~\cite{metz}, or starting directly from the definition of the density of state $N \rho_N (E) = \lim_{\eta \to 0^+} {\rm Tr} \, {\rm Im} {\cal G}$.
Here we follow the second path, and write:
\[
	N \rho_N (E) = \frac{1}{\pi} \lim_{\eta \to 0^+} \frac{{\rm Im} \sum_{i=1}^N \int {\cal D} \phi \, \phi_i^2 \, 
	e^{-\frac{1}{2} \sum_{j,k} \phi_j ({\cal H} - z)_{jk} \phi_k}}{ Z(z)} \, ,
\]
where the ``partition function'' $Z(z)$ is defined as:
\[
	Z(z) = \int {\cal D} \phi \, e^{-\frac{1}{2} \sum_{j,k} \phi_j ({\cal H} - z)_{jk} \phi_k} = \frac{\pi^{N/2}}{\sqrt{{\rm det} ({\cal H} - z)}} \, ,
\]
and $z = E + i \eta$.
From the expressions above, it is straightforward to rewrite the DoS as:
\[
	N \rho_N (E) = \frac{1}{i \pi} \lim_{\eta \to 0^+} \left[ \frac{\partial \ln Z(z)}{\partial E} - \frac{\partial \ln Z(z^\star)}{\partial E} \right]
	\, .
\]
Inserting this equation into the definition of the number of energy levels within the interval $[-E/2,E/2]$, ${\cal N}_N (E) = N \int_{-E/2}^{E/2} 
\rho_N (E^\prime) {\rm d} E^\prime$, one finally ends up with:
\begin{equation} \label{eq:NE}
	\begin{aligned}
		{\cal N}_N (E) &= \frac{1}{i \pi} \lim_{\eta \to 0^+} \big[ \ln Z(E/2+i \eta) - \ln Z(-E/2+i \eta) \\
		& \qquad \qquad - \ln Z(E/2-i \eta) + \ln Z (-E/2 - i \eta)
	\big] \, .
	\end{aligned}
\end{equation}
The ``generalized free-energy'' $F(z) = \ln Z(z)$ can be easily computed within the BP approach as a sum of local contributions involving the
Green's functions defined on the nodes of the RRG and the cavity Green's functions defined on the links of the RRG. 
More precisely it can be shown that $\ln Z(z)$ can be written as a sum of a site and a link contributions~\cite{mezard,PopDyn}:
\begin{equation} \label{eq:F1}
	F(z) = \sum_{i=1}^N \Delta F_s^{(i)} (z) - \sum_{\langle i,j \rangle} \Delta F_l^{(i \leftrightarrow j)} (z) \, ,
\end{equation}
where $\Delta F_s^{(i)} (z)$ is the ``free-energy shift'' corresponding to the addition of site $i$ to the lattice:
\[
\begin{aligned}
	e^{\Delta F_s^{(i)}} & = \frac{\int {\rm d} \phi_i \prod_{j} {\rm d} \phi_j \, e^{ (\epsilon_i + z) \frac{\phi_i^2}{2} - 
	\sum_{j} \left[ \frac{\phi_j^2}{2 G_{j \to i}} - t \phi_i \phi_j \right]}}
	{\int \prod_{j} {\rm d} \phi_j \, e^{- \sum_{j} \frac{\phi_j^2}{2 G_{j \to i}}}}  \\
	& = \sqrt{2 \pi {\cal G}_i} \, ,
	\end{aligned}
\]
where the index $j$ runs over the $k+1$ neighbors of $i$,
and $\Delta F_l^{(i \leftrightarrow j)} (z)$ is the ``free-energy shift'' corresponding to the addition of the link between sites $i$ and $j$:
\[
\begin{aligned}
	e^{\Delta F_l^{(i \leftrightarrow j)}} & = \frac{\int {\rm d} \phi_i \, {\rm d} \phi_j \, e^{- \frac{\phi_i^2}{2 G_{i \to j}} - \frac{\phi_j^2}{2 G_{j \to i}} + t \phi_i \phi_j}}
	{\int {\rm d} \phi_i \, {\rm d} \phi_j \, e^{- \frac{\phi_i^2}{2 G_{i \to j}} - \frac{\phi_i^2}{2 G_{i \to j}}}} \\
        & = \left(1 - t^2 G_{i \to j} G_{j \to i}\right)^{-1/2} \, .
        \end{aligned}
\]
In fact, the addition of a site $i$ can be equivalently viewed as a two-step process: first 
the cavity iteration involving the site $i$ and only $k$ of its $k+1$ neighbors (say, sites $\{ j_1, \ldots, j_k \}$) and then the addition
of the link between the cavity site $i$ and the missing neighbors $j_{k+1}$. Hence one has that~\cite{mezard,PopDyn};
\[
	\Delta F_s^{(i)} (z) = \Delta F_{\rm iter}^{(i \to j)} (z) + \Delta F_l^{(i \leftrightarrow j)} (z) \, ,
\]
which implies that the ``free-energy'' (\ref{eq:F1}) can be equivalently rewritten as:
\begin{equation} \label{eq:F2}
	F(z) = \frac{1-k}{2} \sum_{i=1}^N \Delta F_s^{(i)} (z) + \sum_{\langle i,j \rangle} \Delta F_{\rm iter}^{(i \to j)} (z) \, ,
\end{equation}
where the ``iteration free-energy shift'' reads:
\[
	e^{\Delta F_{\rm iter}^{(i \to j)}} = \sqrt{2 \pi G_{i \to j}} \, .
\]
Plugging the ``free-energy shifts'' into Eqs.~(\ref{eq:F1}) and (\ref{eq:F2})
one finds two equivalent expressions for the generalized free-energy:
\begin{equation} \label{eq:F}
\begin{aligned}
F(z) & = \frac{1}{2} \sum_{i=1}^N \ln [ 2 \pi {\cal G}_i (z) ] + \frac{1}{2} \sum_{\langle i, j \rangle} \ln [ 1 - t^2 G_{i \to j} (z) G_{j \to i} (z) ] \\
& = 
\frac{1-k}{4} \sum_{i=1}^N \ln [ {\cal G}_i (z) ] + \frac{1}{2} \sum_{\langle i, j \rangle} \ln [G_{i \to j} (z)] + \frac{N}{2} \ln (2 \pi) \, ,
\end{aligned}
\end{equation}
Using the iteration equations~(\ref{eq:recursion}) and (\ref{eq:recursion_final}), by noticing that ${\cal G}_i^{-1} = G_{i \to j}^{-1} - t^2 G_{j \to i}$, 
one can explicitly show that these two expressions are in fact the same.
Furthermore, since ${\cal G}_i (z^\star) = {\cal G}_i^\star (z)$ one has that:
\[
\ln \frac{{\cal G}_i (z)}{ {\cal G}_i (z^\star)} = 2 i \Psi_i (z) \, ,
\]
where ${\cal G}_i (z) = | {\cal G}_i (z) | e^{i \Psi_i (z)}$. (From now on we choose to define the angles in the interval $[-\pi,\pi]$, i.e., we place
the branch-cut of the logarithm along the negative real axis. In fact, since the imaginary part of the Green's functions are all positive for $\eta > 0$ by definition, 
all the $\Psi_i$ and $\psi_{i \to j}$ involved in the equations will fall in the interval $[0,\pi]$.)
Hence, plugging the second line of Eq.~(\ref{eq:F}) into Eq.~(\ref{eq:NE}) one finds Eq.~(\ref{eq:NE1}) given in the main text.
Equivalently, from the first line of Eq.~(\ref{eq:F}) one gets:
\begin{equation} \label{eq:NE2}
\begin{aligned}
{\cal N}_N (E) =& \frac{1}{\pi} \! \lim_{\eta \to 0^+} \! \bigg \{ \sum_{i=1}^N \big[ \Psi_i ( z_+ ) - \Psi_i (  z_- ) \big]\\
& \qquad \,\,\,\,\,\, + \sum_{\langle i, j \rangle} \big[ \varphi_{i \leftrightarrow j} ( z_+ ) - \varphi_{i \leftrightarrow j} (  z_- ) \big] \bigg \} \, ,
\end{aligned}
\end{equation}
where $z_{\pm} = \pm E/2 + i \eta$, and the angle $\varphi_{i \leftrightarrow j} ( z)$ is defined as the phase of $1 - t^2 G_{i \to j} (z) G_{j \to i} (z)$.

For a random diagonal Hamiltonian, ${\cal H} = - \epsilon_i \delta_{ij}$ (i.e., $t=0$), for which one has that ${\cal G}_i = G_{i \to j} = (- \epsilon_i - z)^{-1}$, 
one can explicitly check using the representation of the Heaviside step function in terms of the discontinuity 
of the complex logarithm along the negative real axis that both Eqs.~(\ref{eq:NE1}) and (\ref{eq:NE2}) both give back
${\cal N}_N (E) = \int_{-E/2}^{E/2} \delta( E^\prime + \epsilon_i) {\rm d} E^\prime$.

The computation of $F(z)$ on the Cayley tree is even easier, 
since one can obtain its expression directly by integrating
out progressively the sites starting from the boundary.
This yields:
\[
	Z(z) = \left ( 
	\prod_{\ell = 1}^{n_g} \prod_{i_\ell = 1}^{N_\ell} \sqrt{2 \pi G_{i_\ell \to i_{\ell-1}} (z)}
	\right) \sqrt{2 \pi {\cal G}_0 (z)} \, ,
\]
where $N_\ell = (k+1) k^{\ell - 1}$ is the total number of sites $i_\ell$ belonging to the $\ell$-th
generation of the tree. Plugging this expression into Eq.~(\ref{eq:NE}) one finally obtains:
\begin{equation} \label{eq:NECT}
\begin{aligned}
	{\cal N}_N (E) =& \frac{1}{\pi} \! \lim_{\eta \to 0^+} \! \bigg \{ \Psi_0 ( z_+ ) - \Psi_0 (  z_- )\\
	& \,\,	
	\sum_{\ell = 1}^{n_g} 
	\sum_{i_{\ell}=1}^{N_\ell} \big[ \psi_{i_\ell \to i_{\ell-1}} ( z_+ ) - \psi_{i_\ell \to i_{\ell-1}} 
	(  z_- ) \big] \bigg \} \, ,
\end{aligned}
\end{equation}

\begin{acknowledgments}
We thank I. Aleiner, B. L. Altshuler, E. Bogomolny, J.-P. Bouchaud, C. Castellani, Y. Fyodorov, T. Garel, L. Ioffe, 
	V. Kravtsov, P. Le Doussal, G. Lemari\'e, A. D. Mirlin, 
C. Monthus, M. Muller, V. Oganesyan, G. Parisi, V. Ros, A. Scardicchio, G. Semerjian, K. S. Tikhonov, S. Warzel for useful inputs, remarks and discussions. This research was partially supported by a grant from the Simons Foundation ( \# 454935 Giulio Biroli). Marco Tarzia is a member of the Institut Universitaire de France.
\end{acknowledgments}



\begin{thebibliography}{99}
\bibitem{precisazioneCT} Note that on the Cayley tree
	the number of coupled equations for the cavity Green's functions are $2 (N-1)$ instead of $(k+1) N$,	
	due to the fact that the boundary sites only have one neighbors.
\end{thebibliography}
\end{document}